\documentclass[aps, prd, nofootinbib, superscriptaddress, preprint]{revtex4}
\usepackage{amsmath}%,amssymb}
\usepackage{amssymb}
\usepackage{makeidx}
\usepackage{gensymb}
\usepackage{amsfonts}
\usepackage[ansinew]{inputenc} % font encodings
\usepackage[usenames,dvipsnames]{pstricks}
\usepackage{pst-grad} % For gradients
\usepackage{graphicx}
\usepackage{caption}
\usepackage{color}
\usepackage{allpurpose}

\begin{document}

\title{Escape, bound and capture geodesics in local static coordinates in Schwarzschild spacetime}

\author{Yaoguang Wang}
\affiliation{School of Physics and Technology, Wuhan University, Wuhan, 430072, China}

\author{Xionghui Liu}
\affiliation{Quantum Information Research Center, Shangrao Normal University, Shangrao, Jiangxi 334001, China}

\author{Nan Yang}
\affiliation{Glyn O. Phillips Hydrocolloid Research Centre,  Hubei University of Technology, Wuhan 430068, China}
\email{nanyang27@gmail.com}

\author{Jiawei Liu}
\affiliation{School of Physics and Technology, Wuhan University, Wuhan, 430072, China}

\author{Junji Jia}
\affiliation{Center for Astrophysics, School of Physics and Technology, School of Physics and Technology, Wuhan University, Wuhan, 430072, China}
\affiliation{MOE Key Laboratory of Artificial Micro- and Nano-structures, School of Physics and Technology, School of Physics and Technology, Wuhan University, Wuhan, 430072, China}
\email{junjijia@whu.edu.cn}

\date{\today}

\begin{abstract}

The classical geodesics of timelike particles in Schwarzschild spacetime is analyzed according to the particle starting radius $r$, velocity $v$ and angle $\alpha$ against the radial outward direction in the reference system of an local static observer. The region of escape, bound and capture orbits in the parameter space of $(r,~v,~\alpha)$ are solved using the three cases of the effective potential. It is found that generally for radius smaller than $4M$ or velocity larger than $c/\sqrt{2}$ there will be no bound orbits. While for fixed radius larger than $4M$ (or velocity smaller than $c/\sqrt{2}$), as velocity (or radius) increase from zero (or $2M$), the particle is always captured until a critical value $v_{\mathrm{crit1}}$ (or $r_{\mathrm{crit1}}$) when the bound orbit start to appear around $\alpha=\pi/2$ between a double-napped cone structure. As the velocity (or radius) increases to another critical value $v_{\mathrm{crit2}}$ (or $r_{\mathrm{crit2}}$) then the bound directions and escape directions in the outward cone become escape directions, leaving only the inward cone separating the capture and bound directions. The angle of this cone will increase to its asymptotic value as velocity (or radius) increases to its asymptotic value. The implication of these results in shadow of black holes formed by massive particles, in black hole accretion and in spacecraft navigation is briefly discussed.

\end{abstract}

\keywords{Schwarzschild spacetime, escape cone, bound orbits, shadow}

\maketitle

\section{Introduction}

The study of test particle geodesics in a spacetime in General Relativity is important because the knowledge of the spacetime can be completely deduced from the geodesic motions in it and vice versa. Among known spacetimes, Schwarzschild spacetime is one of the simplest and consequently numerous studies on geodesic motions in it exist in literature, emphasizing different aspects of theoretical considerations or astrophysical applications. A very incomplete collection of these lies in Ref. \cite{Suzuki:1996gm,DOrazio:2010nbl,Scharf:2011ii,Barack:2011ed,Gibbons:2011rh,DeFalco:2016yox,Tejeda:2013mva,Tsupko:2014wza,Gorbatenko:2015ing}.

Among motion of different kinds of particles in Schwarzschild and related spacetimes, the motion of photons is the most intensively studied (see \cite{hagi,darwin,mielnik,metzner,zeldovich,atkinson,synge1} for early works). In particular, Synge
found that photons can only escape to infinity in the Schwarzschild spacetime if their escaping direction was within a cone region (see Ref. \cite{synge1} and Fig. \ref{mlessparaspace}). This cone was called the {\it escape cone} in later literature and is the complement of the {\it cone of avoidance} in Chandrasekhar's book \cite{Chandrasekhar:1985kt}. This is now considered as a pioneering work in the study of shadow of black holes (BH) \cite{Perlick:2004tq,Grenzebach:2014fha,Grenzebach:2015oea,Perlick:2015vta,Abdujabbarov:2016hnw}, which is recently observed for the M87 central BH \cite{Akiyama:2019cqa, Akiyama:2019eap}.
The timelike geodesics for massive test particles in Schwarzschild spacetime was studied in Ref. \cite{hagi, mielnik, darwin, metzner} and classified in Ref. \cite{Misner:1974qy,Chandrasekhar:1985kt} in terms of conserved specific orbital angular momentum $L$ and specific energy $E$ of the test particle. However, $L$ and $E$ of a test particle can not be conveniently determined locally, i.e., they are more easily measured by external observers far away but not a local static observer. The quantities that are natural in the local static reference system and can be connected with $E$ and $L$ are the local radial coordinate $r$, the local particle velocity $v$ and the velocity direction. In this work, we would like to focus on the following questions: suppose that a timelike test particle is located in Schwarzschild spacetime at radius $r$, with local velocity $v$ and velocity direction angle $\alpha$ against the outward radial direction, then what kind of geodesic will this test particle have, i.e., will the particle fall into the BH, move along a bound orbit or escape to infinity.

We will show that for massive particles starting from radius $r$ and with local velocity $v$, there will be no bound orbit but only escape or capture ones if the radius is smaller than $4M$ or the velocity is larger than $c/\sqrt{2}$. If the radius is larger than $4M$, then bound orbits will appear when the velocity is larger than a critical value $v_{\mathrm{crit1}}$ and all these bound orbits should have initial velocities pointing outside a double-napped cone structure (see Fig. \ref{figfixedrchangev} (a)). Directions inside the cones lead to capture orbits. The angle of the inward cone against the radial outward direction will increase as the velocity further increase until a second critical value $v_{\mathrm{crit2}}$ beyond which the bound directions and outward capture directions become escape directions. As the velocity further increases, the angle of the inward cone separating the capture and escape directions will further increase until its asymptotic value prescribed by a simple formula for massless particles. The process is similar if one fixes the velocity $v$ to a value smaller than $c/\sqrt{2}$ but increases the starting radius $r$.
We will solve the exact formula for the angle of the cones $\alpha_e(r,v)$ of massive test particle as a function of radius $r$ and velocity $v$. Moreover, all the critical value of velocity and radius will also be obtained.

The work is organized as follows. In section \ref{setup} we setup the radial geodesic equation in the Schwarzschild metric and summarize pervious result in the null case. In section \ref{sec:mpmotion} we do the case study of the massive particle motion and partition the parameter space of $(r,~v,~\alpha)$ into the regions of escape, bound and capture geodesics. Lastly in section \ref{sec:impdis}
we discuss the implication of this work and its possible extensions.

\section{The setup and motion of massless particles\label{setup}}

The Schwarzschild metric takes the form
\begin{align}
\dd s^2=-\left(1-\frac{2M}{r}\right)\dd t^2+\left(1-\frac{2M}{r}\right)^{-1}\dd r^2+r^2(\dd\theta^2+ \sin^2\theta \dd\phi^2),
\label{metric}
\end{align}
in which $(t,r,\theta,\phi)$ are the coordinates and $M$ is the mass of the BH.
The geodesic equation together with initial conditions completely determines the orbits of test particles in this spacetime. We are only interested in the final state of particle's motion, i.e. being captured, bounded or escape to infinity, but not the detailed shape of the orbit. Therefore we merely need to focus on the radial geodesic equation, which is
\be
\frac{1}{2}\left(\frac{\dd r}{\dd\tau}\right)^2+V =\frac{1}{2}E^2,\label{req}
\ee
where $\tau$ is the affine parameter and $V$ is the effective potential given by
\be
V =\frac{1}{2}\left(1-\frac{2M}{r}\right) \left(\frac{L^2}{r^2}+\kappa\right).
\label{veffdef}
\ee
Here $\kappa=0,~1$ for massless and massive particles respectively. $E$ and $L$ are two constants of motion introduced in the first integration of the time and angular geodesics equations
\bea
E&=&\left(1-\frac{2}{r}\right)\frac{\dd t}{\dd \tau},\label{edef}\\
L&=& r^2\frac{\dd\phi}{\dd\tau}.\label{ldef}
\eea
Note that in Schwarzschild spacetime, the geodesic motions are planar and therefore we can always set $\theta(\tau)=\pi/2$. $E$ and $L$ are interpreted respectively as the specific energy of the test particle at infinite radius and orbital angular momentum of the test particle around the $\theta=0$ axis.

Eq. \refer{req} is mathematically equivalent to the equation describing a one-dimensional classical mechanical motion of a particle with unit mass and total energy $\displaystyle E^2/2$ in an effective potential $V$. Consequently, the relationship between $\displaystyle E^2/2$ and $V$ will determine the final state of the particle. Because we are using the natural units $G=c=1$, the length and time will have the same dimension as mass. Therefore to simplify the analysis, we will perform and use henceforth the following change of variables
\begin{align}
\frac{r}{M}\to r, ~\frac{\tau}{M}\to \tau, ~\frac{L}{M}\to L. \label{chvar}
\end{align}
With this change, the effective potential \eqref{veffdef} becomes
\begin{align}
V=\frac{1}{2}\left(1-\frac2{r}\right)\left(\frac{L^2}{r^2}+\kappa\right).
\label{veff}
\end{align}

\begin{figure}[htp!]
\includegraphics[width=0.45\textwidth]{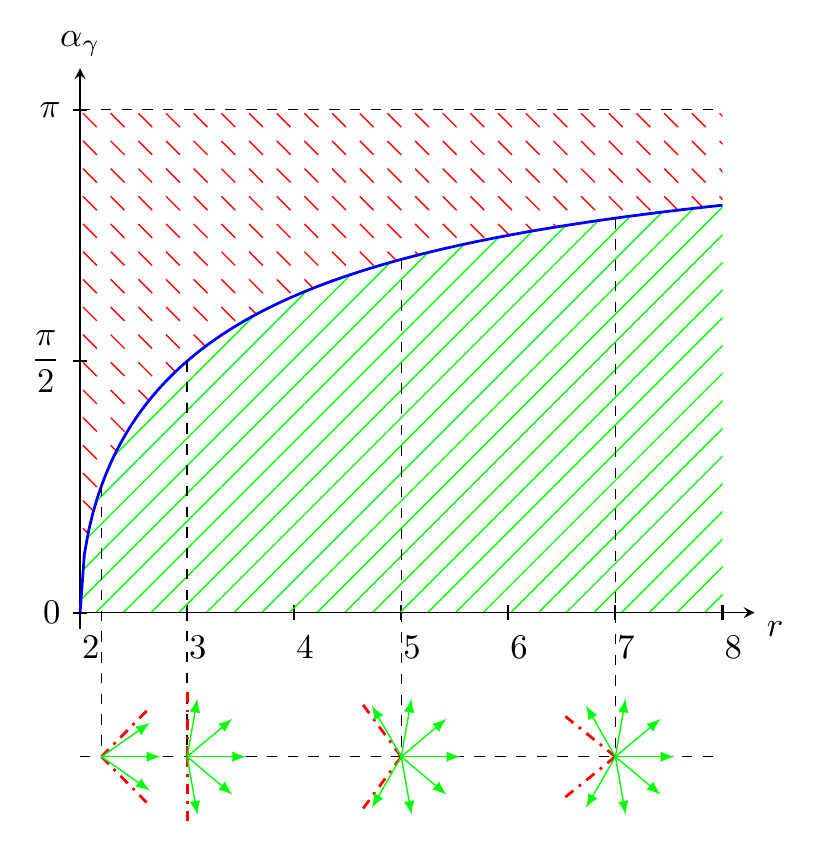}
\caption{The parameter space for photons that allows escape (solid green region) and capture (dashed red region) orbits. The dotdashed cones under $x$ axis are the {\it escape cones} at different radii.  \label{mlessparaspace}}
\end{figure}

The final states for the  motion of massless particles starting from radius $r$ with local velocity $c$ and velocity angle $\alpha$ against the radial outward direction has been studied in Ref. \cite{synge1,Chandrasekhar:1985kt}. For completeness, we summarize their findings here. The most essential result in this case is a critical value of the velocity angle against the radial outward direction, which is  given by
\be
\alpha_\gamma(r)=\arccos\left(-\frac{(r-3) \sqrt{r+6}}{r^{3/2}}\right)\label{photoec}
\ee
(see Fig. \ref{mlessparaspace}).
This angle is half of the local opening of an escape cone only within which massless particle can escape. In other words, it divides  the 2-dimensional parameter space spanned by $(r,~\alpha)$  into two regions: the region of escape orbits and the region of capture orbits. Massless particles starting in the escape (or capture) region will escape from (or be captured by) the hole to infinity (or its singularity). As illustrated in Fig. \ref{mlessparaspace}, the green arrows below $x$ axis denote the escape directions and the dotdashed red boundaries denote the critical cone angle $\alpha_\gamma(r)$ at the corresponding radius.

\section{Motion of massive particles\label{sec:mpmotion}}

For massive particles, the parameter space has one more dimension, the particle velocity $v$, and therefore becomes $(r,~v,~\alpha)$.
Depending on the point in this space, which corresponds to the initial conditions of the particle, the effective potential $V(r,L)$ that the test particle experience will be different. Consequently, unlike the case of photons, there is a possibility of a bound orbit  in addition to the escape and capture geodesics.

In order to clarify which part of the parameter space will lead to respectively escape, bound and capture orbits, we first analyze the effective potential \refer{veffdef} and then connect its properties to different regions in the parameter space. Using $\kappa=1$ in Eq. \refer{veffdef}, the potential, now denoted by $V_1$, becomes
\begin{align}
V_1=\frac{1}{2}\left(1-\frac2{r}\right)\left(\frac{L^2}{r^2}+1\right).
\label{veff1}
\end{align}
This potential equals 0 at the surface of the BH and $\frac{1}{2}$ at infinity. In between, the potential can have local extrema if the extremal condition
\be \frac{\partial V_1(r,L)}{\partial r}=0 \label{k1econd}\ee
has real solutions of $r$ that are larger than 2.

\begin{figure}[htp!]
\includegraphics[width=0.3\textwidth]{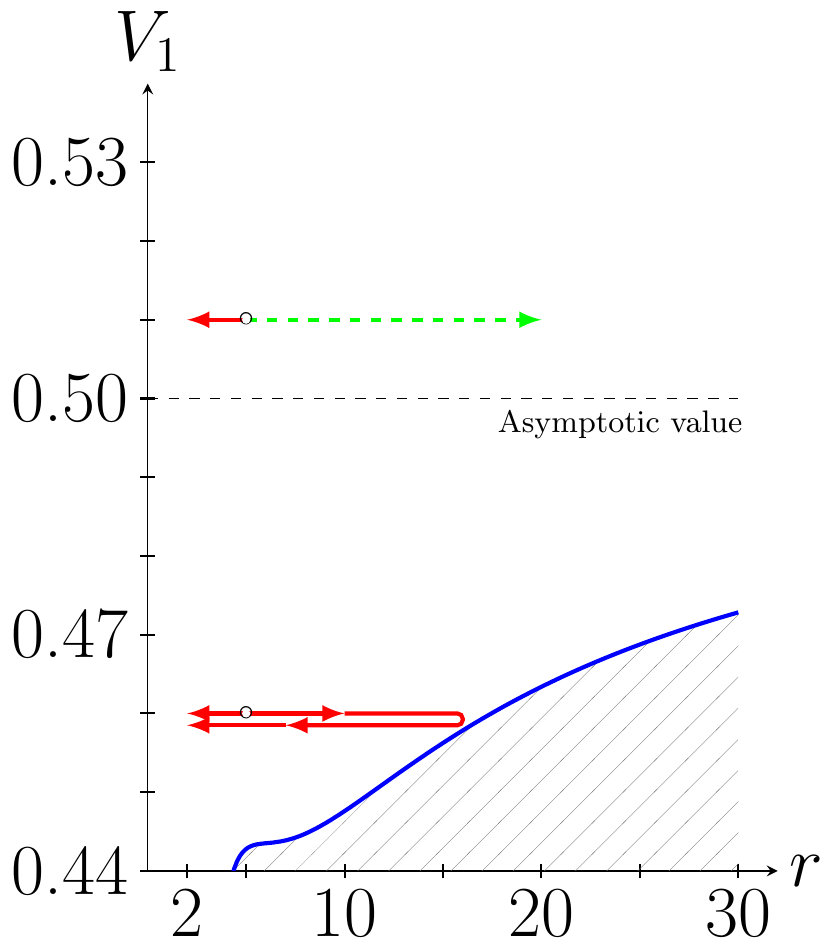}
\includegraphics[width=0.3\textwidth]{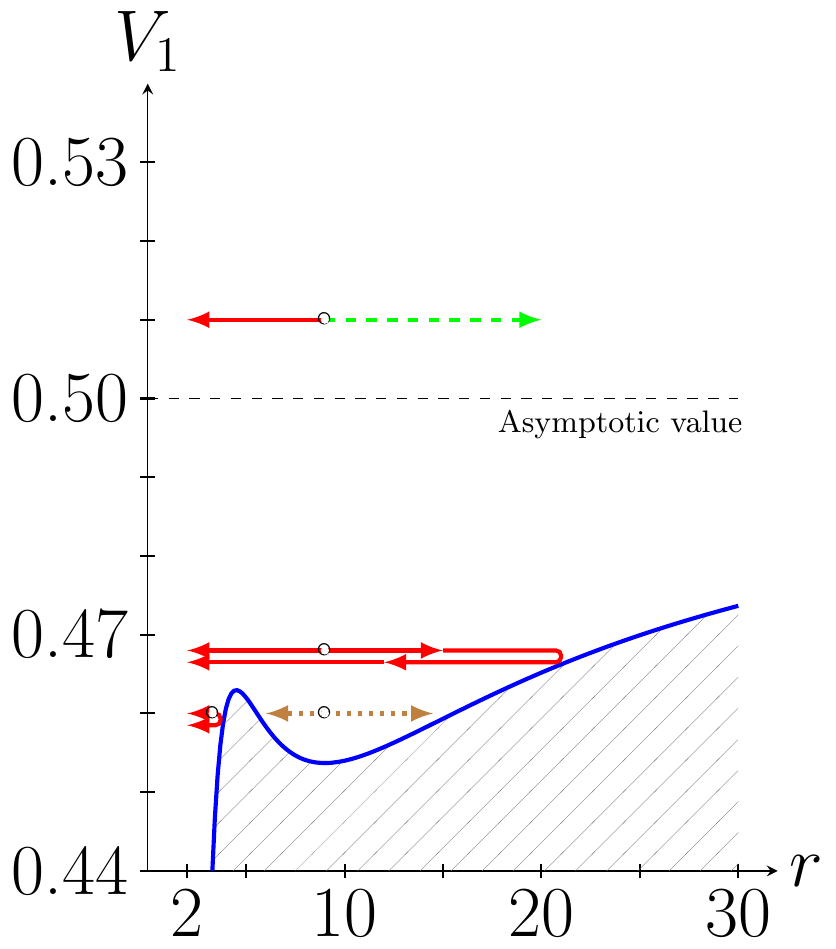}
\includegraphics[width=0.3\textwidth]{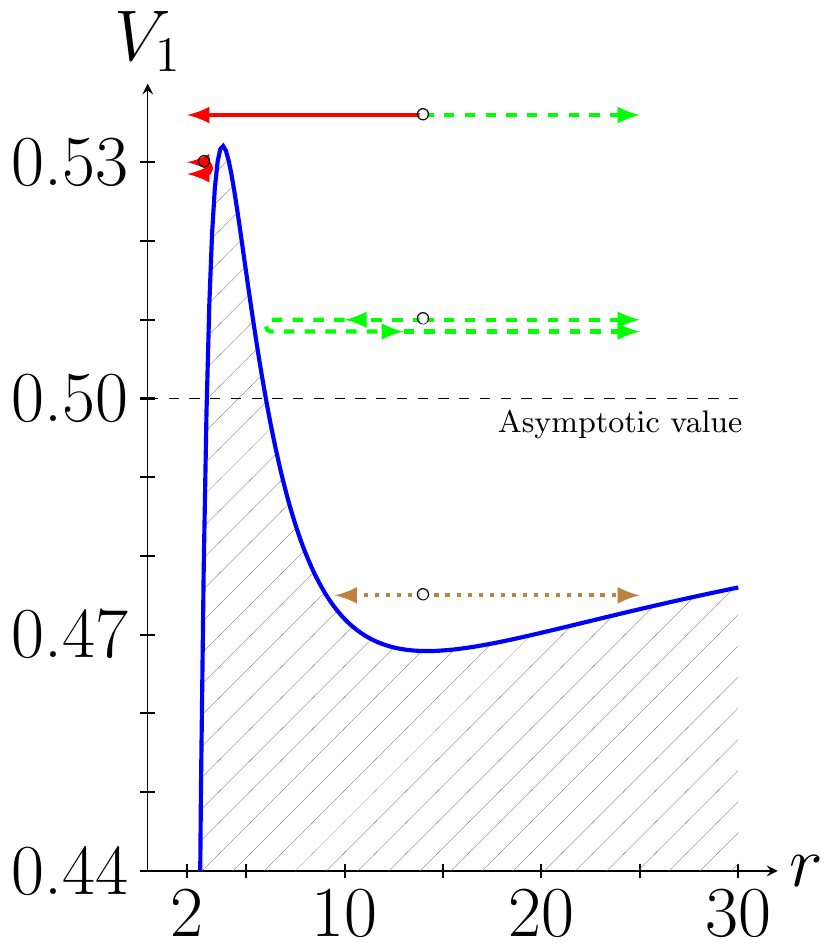}\\
(a)\hspace{4cm}(b)\hspace{5cm}(c)
\caption{The effective potential as a function of radius for some fixed $L$: (a) $L^2=11.9$, (b) $L^2=13.5$ and (c) $L^2=18$. In each case, typical escape (dashed green), bound (dotted brown) and capture (solid red) orbits are indicated.  \label{figpot}}
\end{figure}

Solving Eq. \refer{k1econd}, we find
\begin{align}
r_\pm=\frac{1}{2}\left(L\pm\sqrt{L^2-12}\right)L. \label{rpmdef}
\end{align}
Thus if $L^2<12$, there is no real solutions and no local extreme in $r\in (2,\infty)$. In this case, the potential $V_1$ increases monotonically from 0 to its asymptotic value $\frac{1}{2}$.
When $L^2=12$, the two roots $r_\pm$ are degenerate
\be r_+=r_-=6 \ee
and this radius is a flat but non-extremal point of the potential.
For $L^2>12$, it can be readily verified that $r_-$ will be a local maximum and $r_+$ is a local minimum of the potential. Moreover, it is also easy to show that in this case $3<r_-\leq 6$ and $r_+\geq 6$.
Substituting $r=r_+$ and $r_-$ into \refer{veff1} we obtain respectively the maximum and minimum values of the potential
\begin{align}
V_{1,\mathrm{min}}=\frac{\left(L^2+ L\sqrt{L^2-12}-4\right)^2}{L\left(L+ \sqrt {L^2-12}\right)^3},~V_{1,\mathrm{max}}=\frac{\left(L^2- L\sqrt{L^2-12}-4\right)^2}{L\left(L- \sqrt {L^2-12}\right)^3}.
\label{v1m}
\end{align}
One can also verify that the potential minimum $V_{1,\mathrm{min}}$ increases monotonically as $L$ increases but it is always smaller than the asymptotic value $\frac{1}{2}$.
The potential maximum $V_{1,\mathrm{max}}$ also increases with $L$ and will be equal to the asymptotic value when $L^2=16$. Therefore for both the $12\leq L^2<16$ and $L^2\geq 16$ cases, there will be two (local) potential maxima, $V_{\mathrm{1,max}}$ and $1/2$. For the former case, $V_{\mathrm{1,max}}<1/2$ and the particles with $E^2/2$ slightly above the smaller maximum $V_{\mathrm{1,max}}$ will follow a capture geodesic. While for the latter case, the contrary happens, i.e., the particle will escape if its $E^2/2$ is slightly larger than $1/2$ and the initial radius is large. Therefore we can discuss the possible outcomes of the geodesics according to the above three regions of $L^2$.  The corresponding three cases of the potential are summarized in Fig. \ref{figpot} and have appeared in e.g. Ref. \cite{Chandrasekhar:1985kt}.

In the following subsections, we then discuss the possible motion types in each case of the potential in Fig. \ref{figpot}, the connection of the quantities $L$ and $E$ to local observables $v$ and $\alpha$, and most importantly how the physical variables $(r,~v,~\alpha)$ affect the outcome of the orbits.

\bc
\ec

\subsection{Case (1): Small $L^2$ without $V_{\mathrm{1,max}}$ \label{case1}}

In this case, we should require
\begin{align}
L^2<12.  \label{eq:stara}
\end{align}
Then it is seen from Fig. \ref{figpot} (a) that there are only two possibilities for the final state of the geodesics. The particle will escape if and only if $E^2/2$ is equal or higher than the asymptotic value
\be
\frac{1}{2}E^2\geq \frac{1}{2} \label{case1econd}
\ee
and the initial moving direction is not inward
\be
\frac{\dd r}{\dd\tau} \geq 0. \label{case1vdcond}
 \ee
Other particles not satisfying these two conditions in Case (1) will be captured by the BH.

In order to translate the case condition \eqref{eq:stara} and escape conditions \refer{case1econd} and \refer{case1vdcond} to requirements on parameters $r$, $v$ and $\alpha$, we now consider a tetrad $e^\mu_a$ associated with a static observer at coordinates $(r,~\theta,~\phi)$
\bea
e^\mu_0&=&\lb \lb 1-\frac{2}{r}\rb^{-1/2},~0,~0,~0\rb ,\\
e^\mu_1&=&\lb 0,~\lb 1-\frac{2}{r}\rb^{1/2},~0,~0\rb,\\
e^\mu_2&=&\lb 0,~0,~\frac{1}{r},~0\rb,\\
e^\mu_3&=&\lb 0,~0,~0,~\frac{1}{r\sin\theta}\rb.\eea
This tetrad links the Schwarzschild metric \refer{metric} to the Minkovski metric
\be \dd s^2=-\dd \tilde{x}_0^2+\dd \tilde{x}_1^2+\dd \tilde{x}_2^2++\dd \tilde{x}_3^2 \ee
where $(\tilde{x}_0,~\tilde{x}_1,~\tilde{x}_2,~\tilde{x}_3)$ are the coordinates. Note that the spacial direction $\tilde{x}_1$ points to the radial outward direction.
In this local coordinate system, the energy $E$ in Eq. \refer{edef}, angular momentum $L$ in Eq. \refer{ldef} and the $\dd r/\dd \tau$ in Eq. \refer{case1vdcond} become respectively
\bea
E&=&\left(1-\frac{2}{r}\right) \frac{\dd t}{\dd \tau}=\left(1-\frac{2}{r}\right)\frac{\partial t}{\partial \tilde{x}_a}\frac{\dd \tilde{x}_a}{\dd \tau}=\sqrt{1-\frac{2}{r}} \frac{\dd \tilde{x}_0}{\dd \tau} ,\label{edeftrans}\\
L&=& r^2\frac{\dd\phi}{\dd\tau}=r^2 \frac{\partial \phi}{\partial \tilde{x}_a}\frac{\dd \tilde{x}_a}{\dd \tau}
=r
\frac{\dd \tilde{x}_3 }{\dd\tau}=r
\frac{\dd \tilde{x}_3 }{\dd\tilde{x}_0}\frac{\dd\tilde{x}_0}{\dd \tau},\label{ldeftrans}\\
\frac{\dd r}{\dd \tau} &=& \frac{\partial r}{\partial \tilde{x}_a}\frac{\dd \tilde{x}_a}{\dd \tau} =\sqrt{1-\frac{2}{r}} \frac{\dd \tilde{x}_1}{\dd \tau} .\label{vdcondtrans}
\eea
Here we used the fact that $\theta(\tau)=\pi/2$.
In the local system, the quantities in the right hand sides of the above equations are conveniently expressed in terms of the local velocity $v$ of the particle and its angle $\alpha$ against  the radial outward direction
\be
\frac{\dd \tilde{x}_0}{\dd \tau} =\frac{1}{\sqrt{1-v^2}},
~\frac{\dd \tilde{x}_1}{\dd\tilde{x}_0}=v\cos\alpha,~\frac{\dd \tilde{x}_3 }{\dd\tilde{x}_0}=v\sin\alpha. \ee
Using these relations in Eqs. \refer{edeftrans} and \refer{ldeftrans}, we get
\bea
E&=&\frac{\sqrt{1-\frac{2}{r}}}{\sqrt{1-v^2}} ,\label{edeftrans2}\\
L&=&\frac{r v \sin\alpha}{\sqrt{1-v^2}},\label{ldeftrans2}\\
\frac{\dd r}{\dd \tau}&=& \sqrt{1-\frac{2}{r}} v\cos\alpha .\label{vdcondtrans2}
\eea

Substituting Eqs. \refer{edeftrans2}-\refer{vdcondtrans2} into the case condition \eqref{eq:stara}, and the escape conditions \refer{case1econd} and \refer{case1vdcond}, we have
\bea
&&L^2=\frac{r^2 v^2 \sin^2\alpha}{1-v^2}<12, \label{ccondtrans2}\\
&&\frac{1}{2}E^2=\frac{1}{2}\frac{1-\frac{2}{r}}{1-v^2}\geq \frac{1}{2},\label{econdtrans3}\\
&&\frac{\dd r}{\dd \tau} = \sqrt{1-\frac{2}{r}} v\cos\alpha\geq 0 .\label{vdcondtrans3}\eea
Eq. \refer{ccondtrans2}  bound the velocity direction to the range described by
\be
\sin\alpha< \frac{2\sqrt{3}}{r}\sqrt{\frac{1}{v^2}-1}\label{alpha1defsin}.\ee
Condition \refer{econdtrans3} further bounded the escape orbits to the portion in parameter space satisfying
\be v\geq\sqrt{\frac{2}{r}}\label{case1eovres}.\ee
One can show that with $v$ satisfying Eq. \refer{case1eovres}, the right hand side of \refer{alpha1defsin} is less than one, and therefore its inverse sine function can be taken and its solution becomes
\be
\alpha<\arcsin\left(\frac{2\sqrt{3}}{r}\sqrt{\frac{1}{v^2}-1}\right)\equiv\alpha_1 ~\mbox{or}~\alpha>\pi-\alpha_1\label{alpha1def}.\ee
Together with Eq. \refer{vdcondtrans3} which requires $\alpha\leq \pi/2$, then the total requirement of an escape orbit in this case becomes
\be
\alpha<\alpha_1\label{case1eovdres}\ee
and Eq. \eqref{case1eovres}.
The rest of the parameter space in Case (1) then will lead to the capture of the particle.

\begin{figure}[htp!]
\includegraphics[width=0.3\textwidth]{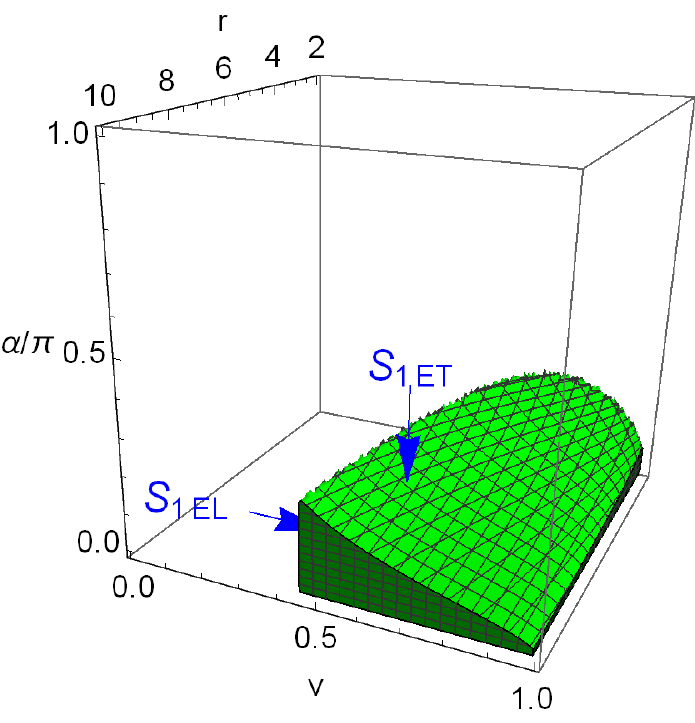}
\includegraphics[width=0.33\textwidth]{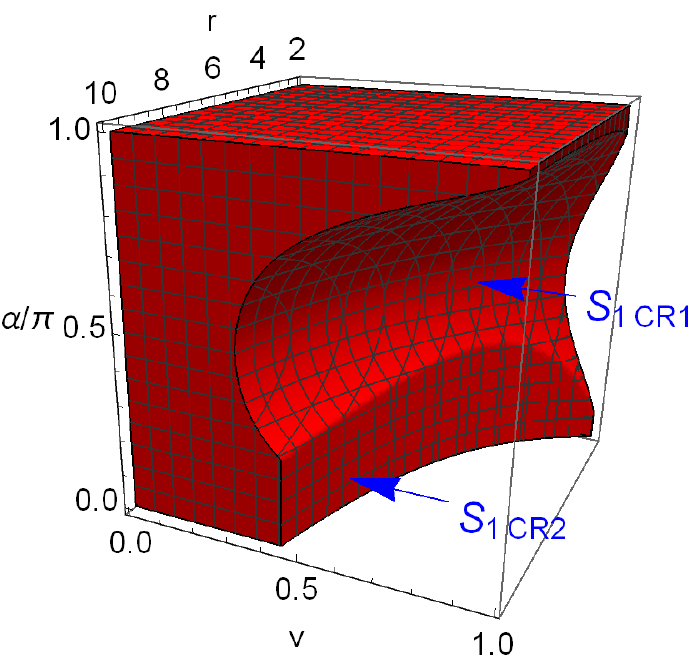}
\caption{Regions in parameter space $(r,~v,~\alpha)$ that allows escape (green) and capture (red) orbits when $L^2<12$. \label{figcase1}}
\end{figure}

Both the regions of escape and capture orbits in Case (1) in the 3-dimensional parameter space $(r,~v,~\alpha)$ are plotted in Fig. \ref{figcase1}. The top surface of the region of escape orbits, labeled as $S_{\mathrm{1ET}}$, and the non-trivial part of the inner boundary of the region of capture orbits, labeled as $S_{\mathrm{1CR1}}$, are plotted using $\alpha_1$ in Eq. \refer{case1eovdres}. The vertical boundary on the back of the region of escape orbits, labeled as $S_{\mathrm{1EL}}$, as well as the vertical boundary on the lower front part of the region of capture orbits, labeled as $S_{\mathrm{1CR2}}$, are plotted using Eq. \refer{case1eovres}.

\subsection{Case (2): Intermediate $L^2$ with $V_{\mathrm{1,max}}<1/2$ \label{case2}}

This case requires
\begin{align}
12\leq L^2< 16. \label{eq:starb}
\end{align}
As stated in the beginning of Sec. \ref{sec:mpmotion}, in this case there exist one local maximum $V_{\mathrm{1,max}}$ at radius $r_-$ and one local minimum $V_{\mathrm{1,min}}$ at radius $r_+$ in the effective potential, as seen in Fig. \ref{figpot} (b) and therefore there exist three possible outcomes of the orbits.
The necessary and sufficient conditions for escape geodesics are formally the same as Eqs. \refer{case1econd} and \refer{case1vdcond}
\be
\frac{1}{2}E^2\geq \frac{1}{2} ,~
\frac{\dd r}{\dd\tau} \geq 0.\label{case2ecs} \ee
The necessary and sufficient conditions for bound orbits are that $E^2/2$ is smaller than the local maximum of the potential \be
\frac{1}{2}E^2< V_{1,\mathrm{max}} \label{case2bdcond}\ee
and the starting radius $r$ is larger than the radius at which the potential is maximal
\be r> r_- .\label{rrel}\ee
Besides the above escape and bound orbits, the particles with initial parameters in other regions allowed by Case (2) will always be captured by the BH. To better understand what values of initial parameters $(r,~v,~\alpha)$ lead to these three kinds of orbits in this case, we now solve the condition \eqref{eq:starb} and \eqref{case2ecs} to \eqref{rrel} in terms of these parameters.

First of all, the condition \eqref{eq:starb} indeed contains two inequalities
\bea
&&L^2\geq 12,\label{lgthan12}\\
\mbox{and}~&&L^2<16.\label{llthan16}
\eea
The first of these, Eq. \refer{lgthan12}, is solved similarly to Eq. \eqref{eq:stara} and the result is similar to Eq. \refer{alpha1def} but with the directions of the inequalities changed
\be
\alpha_1\leq \alpha\leq \pi-\alpha_1,~r>2\sqrt{3\lb\frac{1}{v^2}-1\rb }.\label{case2alphac1}\ee
The second inequality, Eq. \refer{llthan16}, produces a different constraint on $\alpha$
\be
\sin\alpha<  \frac{4}{r}\sqrt{\frac{1}{v^2}-1} .\label{alpha2def2}\ee

Now for the escape orbits, the conditions \refer{case2ecs} should have solutions of the same form as Eq.  \refer{case1eovres} together with the requirement $\alpha<\pi/2$. These solutions guarantee the right hand side of Eq. \refer{alpha2def2} is equal or less than 1 and an inverse sine function can be taken. Therefore, combining with Eqs. \refer{case2alphac1} and \refer{alpha2def2}, the escape condition in Case (2) becomes
\be \alpha_1\leq \alpha<\alpha_2\equiv \arcsin \left( \frac{4}{r}\sqrt{\frac{1}{v^2}-1} \right) , ~ v\geq\sqrt{\frac{2}{r}}.\label{case2ecsol}\ee

For bound orbits, after
substituting Eqs. \refer{v1m}, \refer{edeftrans2} and \refer{ldeftrans2} into  condition \refer{case2bdcond} and taking account into Eq. \refer{lgthan12}, the angle $\alpha$ can be solved to find
\begin{subequations}\label{rrange1}
\begin{align}
\alpha_1<\alpha<\pi-\alpha_1& ~\mbox{when}~\lb 2\sqrt{3\lb \frac{1}{v^2}-1\rb }<r<\frac{18}{1+8v^2},~v>\frac{1}{2}\rb ~\mbox{and} \\
\alpha_e<\alpha<\pi-\alpha_e&~\mbox{when}~\lb r>\frac{18}{1+8v^2}, ~v>\frac{1}{2}\rb~\mbox{or}~\lb r>2\lb \frac{2}{v}-1\rb,~v<\frac{1}{2}\rb,
\end{align}
\end{subequations}
where $\alpha_1$ is given in Eq. \eqref{alpha1def} and  $\alpha_e$ is the angle
\be
\alpha_e\equiv
\arcsin \left(
\sqrt {\frac{8r^2v^4+20r^2v^2-72rv^2-r^2-36r+
108+\sqrt {r-2}\sqrt { \left( 8rv^2+
r-18 \right)^3}}{{2r^3v^2}(rv^2-2)}} \right) .\label{alpha3def}\ee
Using $\alpha_e$, we can define an auxiliary angle satisfying $\sin\alpha_e^\prime =\sin\alpha_e$ but expressed as an inverse cosine function
\begin{align}
\alpha_e^\prime &=\arccos \left(\frac{\left(6-r-\sqrt{r-2} \sqrt{8 r
   v^2+r-18}\right)}{2 \sqrt{2}
   rv}
   \sqrt{\frac{ \sqrt{8 r
   v^2+r-18}(r+2)-\sqrt{r-2}(r-6)}{r \left(\sqrt{8 r
   v^2+r-18}+\sqrt{r-2}\right)}}\right). \label{alphaepdef}
\end{align}
We define this auxiliary $\alpha_e^\prime$ because the $\arcsin$ function in Eq. \eqref{alpha3def} will not yield values in range $(\pi/2,\pi]$ while $\alpha_e^\prime$ can and sometimes it is easier to refer to $\alpha_e^\prime$ than $\alpha_e$ (but not always; otherwise we would have directly used $\alpha_e^\prime$).
When $v=1$, $\alpha_e^\prime$
will reduce to the escape angle $\alpha_\gamma$ of photons in Eq. \refer{photoec}.

The second condition for bound orbits, Eq. \refer{rrel}, when combined with Eq. \refer{lgthan12} has a solution
\begin{subequations}\label{angbd3}
\begin{align}
&\alpha_1<\alpha<\pi-\alpha_1 ~\mbox{when}~\lb r>2\sqrt{3\lb \frac{1}{v^2}-1\rb },~v<\frac{1}{2}\rb  ~\mbox{and}\\
&\alpha_3<\alpha<\pi-\alpha_3~\mbox{when}~\lb r>2+\frac{1}{v^2},~v>\frac{1}{2} \rb,
\end{align}
\end{subequations}
where
\be
\alpha_3\equiv\arcsin\lb \frac{\sqrt{1-v^2}}{v\sqrt{r-3}}\rb
.\ee
One can show that the combination of Eqs. \refer{rrange1} to \refer{angbd3} yields
\be
\alpha_e<\alpha<\pi-\alpha_e,~\mbox{when}~
\lb r>2+\frac{1}{v^2},~v>\frac{1}{2}\rb ~\mbox{or}~
\lb r>2\lb \frac{2}{v}-1\rb,~v<\frac{1}{2} \rb
\ee
and further combination with  Eqs. \refer{case2alphac1} and \refer{alpha2def2} produces the final region of  bound orbits in the parameter space as
\be
\sin\alpha_e<\sin\alpha<\sin\alpha_2~\mbox{and}~\left\{
\begin{array}{ll}
\displaystyle 2\lb \frac{2}{v}-1\rb<r<\frac{2}{v^2}&~\mbox{when}~v<\frac{1}{2} ,~\mbox{and}\\
\displaystyle 2+\frac{1}{v^2}<r<\frac{2}{v^2}&      ~\mbox{when}~\frac{1}{2}<v<\frac{1}{\sqrt{2}}.
\end{array}
\right.\label{case2bcsol}
\ee
Other regions in the parameter space allowed by Case (2) lead to capture orbits.

\bc
\begin{figure}[htp!]
\includegraphics[width=0.3\textwidth]{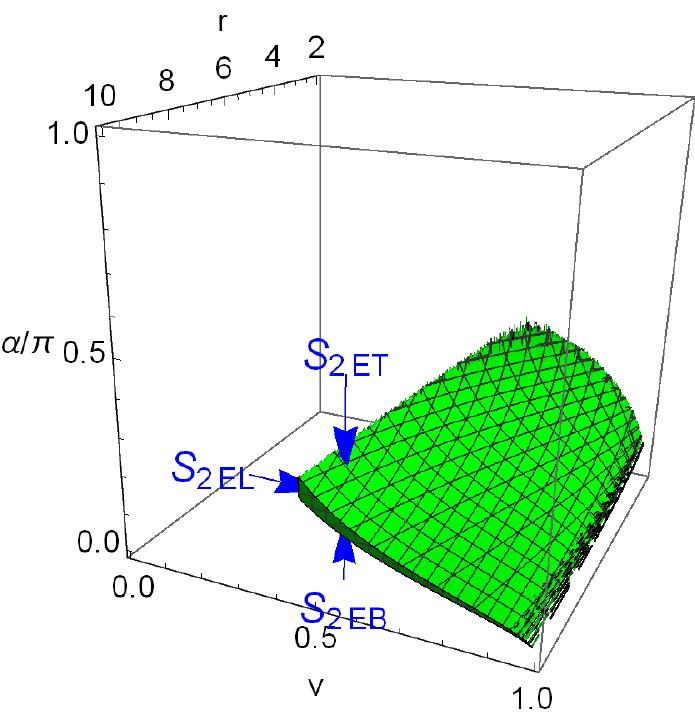}
\includegraphics[width=0.3\textwidth]{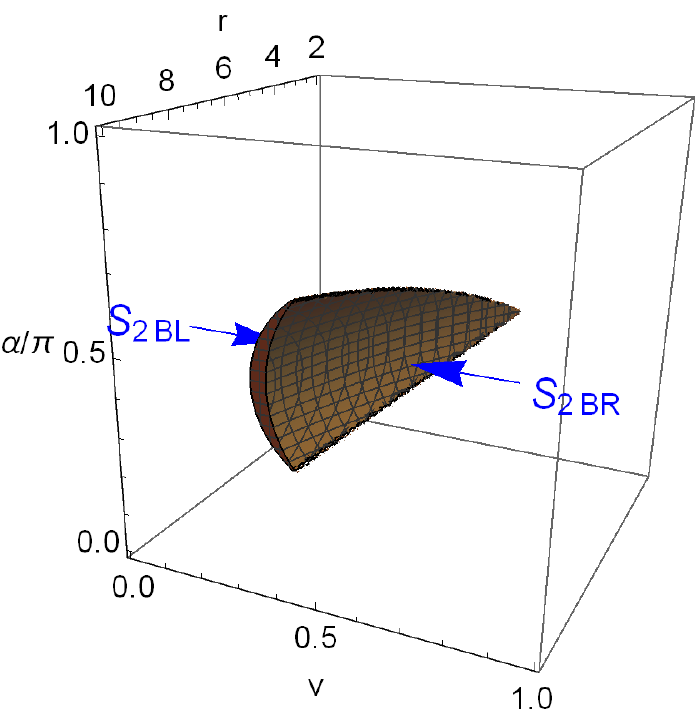}
\includegraphics[width=0.33\textwidth]{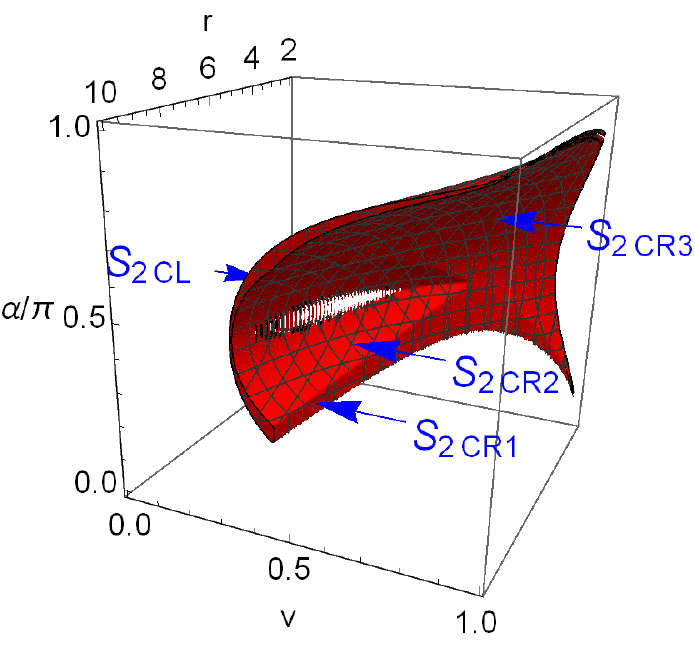}
\caption{Regions in parameter space $(r,~v,~\alpha)$ that allows escape (green),  bound (brown) and capture (red) orbits when $12\leq L^2<16$. The hollow part in the region of capture orbits is a plotting artifact because around that part ($r=6$) the region is infinitely thin. \label{figcase2}}
\end{figure}
\ec

In Fig. \ref{figcase2}, we show all three kinds of regions in the parameter space in Case (2). The top, bottom and backside boundaries of the region of escape orbits, labeled as $S_{\mathrm{2ET}}$, $S_{\mathrm{2EB}}$ and $S_{\mathrm{2EL}}$ respectively, are plotted using $\alpha=\alpha_1$ and $\alpha=\alpha_2$ and relation $v=\sqrt{2/r}$ in Eq. \refer{case2ecsol} respectively. The front and back surface of the region of bound orbits (labeled as $S_{\mathrm{2BR}}$ and $S_{\mathrm{2BL}}$) are plotted using the $\sin\alpha=\sin\alpha_e$ and $\sin\alpha=\sin\alpha_2$ in Eq. \refer{case2bcsol}. The region of capture orbits is also bounded by $\sin\alpha=\sin\alpha_1$ in Eq. \eqref{case2alphac1} (labeled as $S_{\mathrm{2CL}}$) and $\sin\alpha=\sin\alpha_2$ in Eq. \eqref{case2ecsol} (labeled as $S_{\mathrm{2CR3}}$) but with regions of escape and bound orbits removed. After removal, the boundaries  $v=\sqrt{2/r}$ in Eq. \refer{case2ecsol} and $\sin\alpha=\sin\alpha_e$ in Eq. \eqref{case2bcsol} are exposed and labeled as $S_{\mathrm{2CR1}}$ and $S_{\mathrm{2CR2}}$ respectively.

\subsection{Case (3): Large $L^2$ with $V_{\mathrm{1,max}}>1/2$ \label{case3}}

In this case, the requirement on $L^2$ is
\be
L^2\geq 16. \label{eq:starc}
\ee
and the potential and types of geodesics in this case are illustrated in Fig. \refer{figpot} (c). From the derivation of Eq. \refer{alpha2def2}, one sees that the Eq. \eqref{eq:starc} implies
\be
\alpha_2\leq\alpha\leq\pi-\alpha_2.\label{c3alpha2}\ee

The particles in this potential will escape if
\be
\frac{1}{2}E^2>V_{1,\mathrm{max}}, ~ \frac{\dd r}{\dd\tau} \geq 0 \label{case3econd}
\ee
or
\bea
&& \frac{1}{2}\leq \frac{1}{2}E^2\leq V_{1,\mathrm{max}},\label{case3cond2} \\
&& r>r_-. \label{case3rcond2}\eea
and will be bounded if
\bea
&&\frac{1}{2}E^2< \frac{1}{2},\label{case3bdcd1}\\
&&r>r_-. \label{case3bdcd2}\eea
In other regions of the parameter space in Case (3), the particle will be captured.

For the escape orbits, from Eq. \refer{case2bdcond} and its result Eq. \refer{rrange1}, one knows that Eq. \refer{case3econd} should lead to
\be
\alpha_1<\alpha<\alpha_e~\mbox{when}~\lb r>\frac{18}{1+8v^2}, ~v>\frac{1}{2}\rb~\mbox{or}~\lb r>2\lb \frac{2}{v}-1\rb,~v<\frac{1}{2}\rb.\label{c3escape1}\ee
Taking account into Eq. \refer{c3alpha2}, this region of escape orbits is finally constrained by
\be \alpha_2<\alpha<\alpha_e~\mbox{and}~ r>\frac{2}{v^2}.\label{c3escape1f}\ee
For Eq. \refer{case3cond2}, noticing Eqs. \refer{case1econd}, \refer{case2bdcond} and their solutions Eq. \refer{case1eovres}, \refer{rrange1}, its solution is found as
\be
\alpha_e<\alpha<\pi-\alpha_e,~
 r>\frac{2}{v^2}. \label{angbd4}\ee
As in Eq. \refer{rrel}, the condition \eqref{case3rcond2} still yields Eq. \refer{angbd3} which puts on top of Eq.
\refer{angbd4} an extra condition and then for escape orbits satisfying Eq. \eqref{case3cond2}-\eqref{case3rcond2} we have finally
\be
\alpha_e<\alpha<\pi-\alpha_e~\mbox{when}~\lb r\geq2+\frac{1}{v^2},~v>\frac{1}{\sqrt{2}} \rb~\mbox{or}~
\lb r>\frac{2}{v^2},~v<\frac{1}{\sqrt{2}}\rb. \label{angbd5}\ee
The regions in the parameter space bounded by the Eqs. \refer{c3escape1f} and \refer{angbd5}  therefore correspond to the total escape orbits in Case (3).

Now for the region of bound orbits, Eqs. \refer{case3bdcd1} and \refer{case3bdcd2} together with Eq. \refer{c3alpha2} produce
\be  \alpha_2<\alpha<\pi-\alpha_2,~4\sqrt{\frac{1}{v^2}-1}<r<\frac{2}{v^2}, ~v<\frac{1}{\sqrt{2}}.\label{case3bsol}\ee
The rest of the regions in the parameter space in Case (3) will all lead to capture orbits.

\begin{figure}[htp!]
\includegraphics[width=0.32\textwidth]{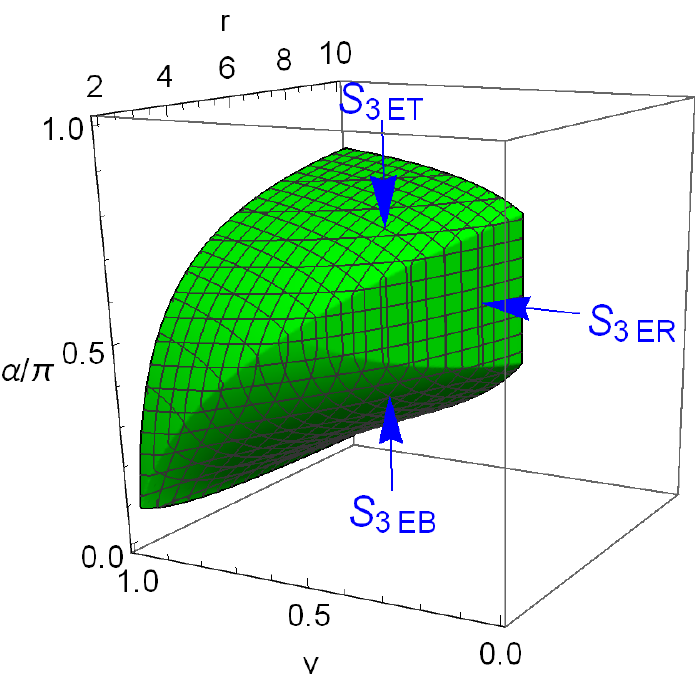}
\includegraphics[width=0.32\textwidth]{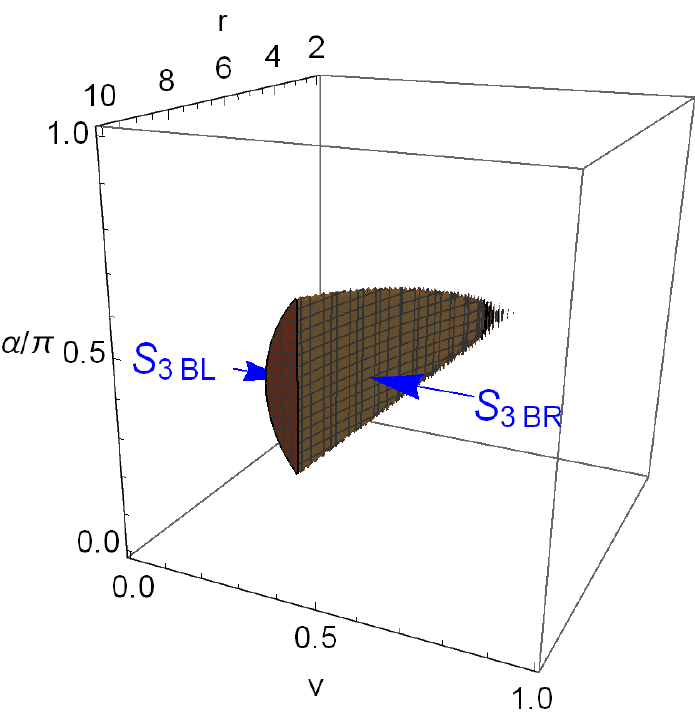}
\includegraphics[width=0.32\textwidth]{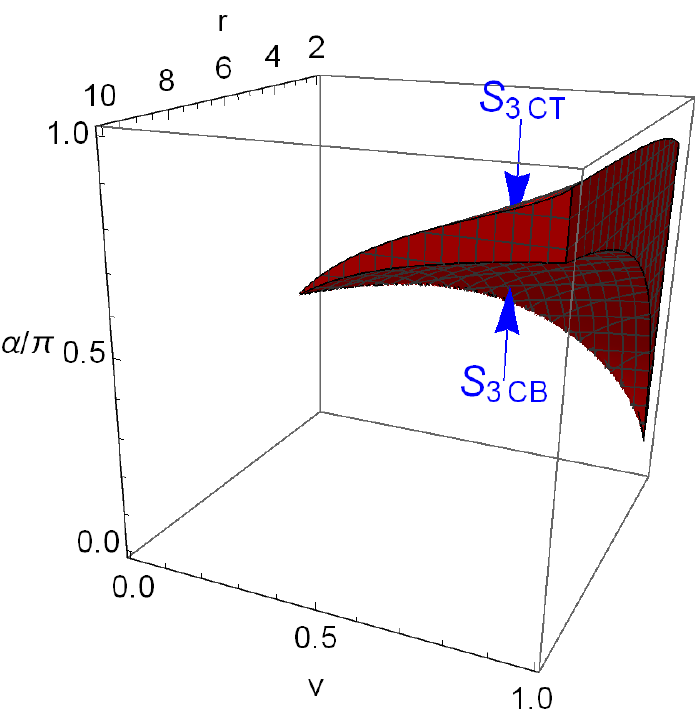}
\caption{The parameter space that allows escape (green), bound (brown) and capture (red) orbits when $L^2\geq 16$. Note that to see the boundary surfaces more clearly, we have rotated the plot of the region of escape orbits 180 degrees around $z$ axis relative to the plots of the regions of bound and capture orbits. \label{figcase3}}
\end{figure}

The regions of escape, bound and capture orbits in the parameter space in this case is shown in Fig. \ref{figcase3}. As seen in Eqs. \refer{c3escape1f} and \refer{angbd5}, the top (labeled as $S_{\mathrm{3ET}}$), bottom (labeled as $S_{\mathrm{3EB}}$) and back (labeled as $S_{\mathrm{3ER}}$) boundaries of the region of escape orbits are described by $\alpha=\alpha_2$, $\alpha=\pi-\alpha_e$ and $r=2/v^2$ respectively. The back and front boundaries (labeled respectively as $S_{\mathrm{3BL}}$ and $S_{\mathrm{3BR}}$) of the region of bound orbits are produced by $\sin\alpha=\sin\alpha_2$ and $r=2/v^2$ respectively, as dictated by Eq. \refer{case3bsol}. Finally, the $\alpha=\pi-\alpha_e$ and $\alpha=\pi-\alpha_2$ are respectively the lower (labeled as $S_{\mathrm{3CB}}$) and upper (labeled as $S_{\mathrm{3CT}}$) surface of the region of capture orbits.

\subsection{Combined regions of escape, bound and capture orbits}

\begin{figure}[htp!]
\includegraphics[width=0.3\textwidth]{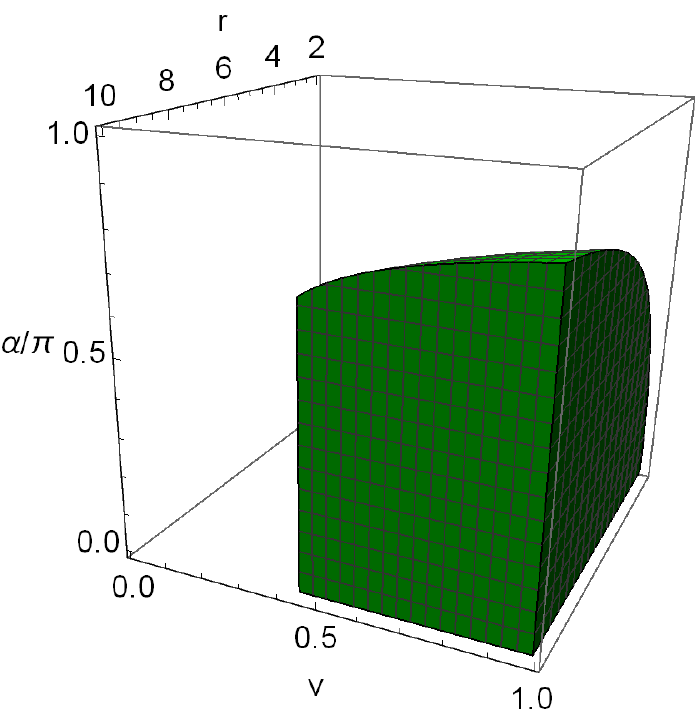}
\includegraphics[width=0.3\textwidth]{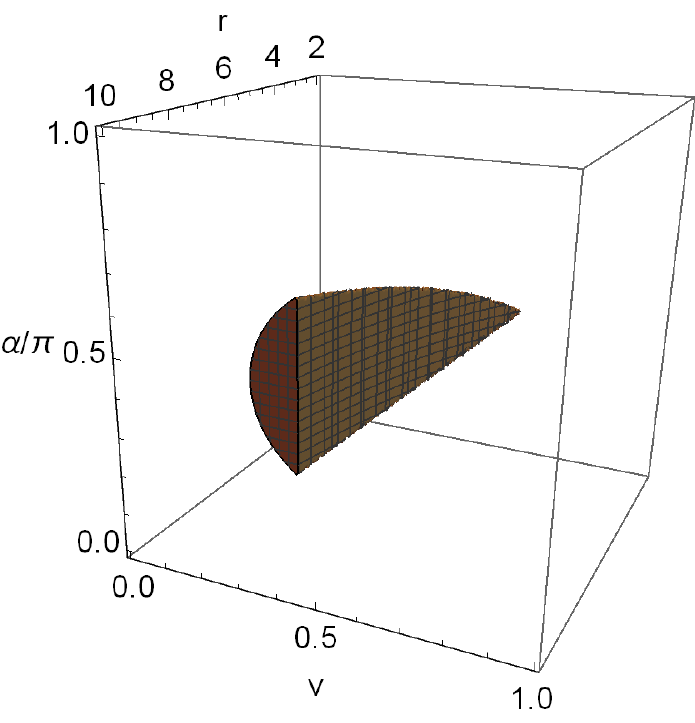}
\includegraphics[width=0.3\textwidth]{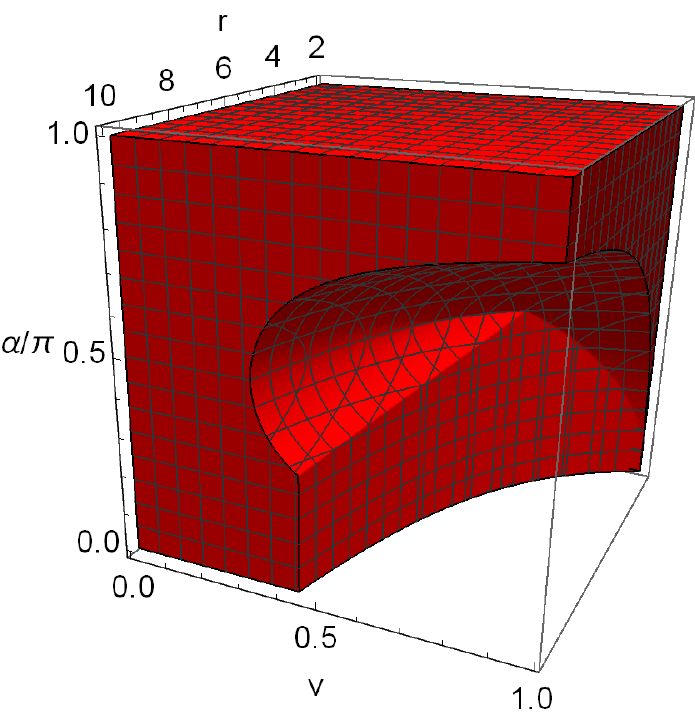}
\caption{The combined regions of escape (green), bound (brown) and capture (red) orbits. These regions completely partition the entire parameter space. Note that to see the boundaries more clearly, we rotated the view angle of the region of escape orbits 180 degrees around $z$ axis relative to those of the regions of bound and capture orbits. \label{figseperate}}
\end{figure}

With all cases solved, we can now combine all regions of escape orbits in Cases (1), (2) and (3). Comparing the labels and origin of the boundaries of the three regions of escape orbits in Figs. \ref{figcase1}, \ref{figcase2} and \ref{figcase3} respectively, one can see that these regions glue together perfectly into one region in the parameter space. Similarly, the combination of all three regions of bound orbits also forms one region and the same happens to all three regions of capture orbits. These results are plotted in Fig. \ref{figseperate}.
It is seen that the region of escape orbits is enclosed by the vertical surfaces $r=2/v^2$, the surface $ \alpha=\pi-\alpha_e$ on the top, as well as the trivial parameter space limits $v=1$ and $\alpha=0$. The region of bound orbits is enclosed by the two surfaces $\sin\alpha=\sin\alpha_e$ and $r=2/v^2$, while the region of capture orbits is bounded by $\sin\alpha=\sin\alpha_e$, $r=2/v^2$ and the trivial parameter space limits $v=0$, $v=1$ and $\alpha=0$, $\alpha=\pi$. Therefore essentially, the three regions in Fig. \ref{figseperate} are separated by two 2-dimensional surfaces $\sin\alpha=\sin\alpha_e$ and $r=2/v^2$. These two surfaces intersect and result in two curves, plotted as the blue and magenta curves respectively in Fig. \ref{figseperate}. Substituting $r=2/v^2$ into $\alpha_e$ in Eq. \eqref{alpha3def} and $\alpha_e^\prime$ in Eq. \eqref{alphaepdef}, one can solve the parametric form of these two curves as $(r,~v,~\alpha)=\lb 2/v^2,~v,~\arccos(2v^2-1)\rb$ with $0<v<1$ for curve one and $(r,~v,~\alpha)=\lb 2/v^2,~v,~\arcsin(2v\sqrt{1-v^2})\rb$ with $v>1/\sqrt{2}$ for curve two. The two curves further intersect at the tip point $(r,~v,~\alpha)=(4,~1/\sqrt{2},~\pi/2)$. This implies that for particles with velocity larger than $1/\sqrt{2}$, there never exists a region of bound orbits. This is indeed expected because the asymptotic energy of particles with such local velocity will be larger than 1 (recalling we are discussing particles with unit mass) and therefore they either escape to infinity or enter the BH but no bound orbit can be formed. While for particles with fixed velocity that is lower than $1/\sqrt{2}$, there can always exist a region in the 2-dimensional parameter space spanned by $(r,~\alpha)$ that allows particles to be bounded.

\bc
\begin{figure}[htp!]
\includegraphics[width=0.24\textwidth]{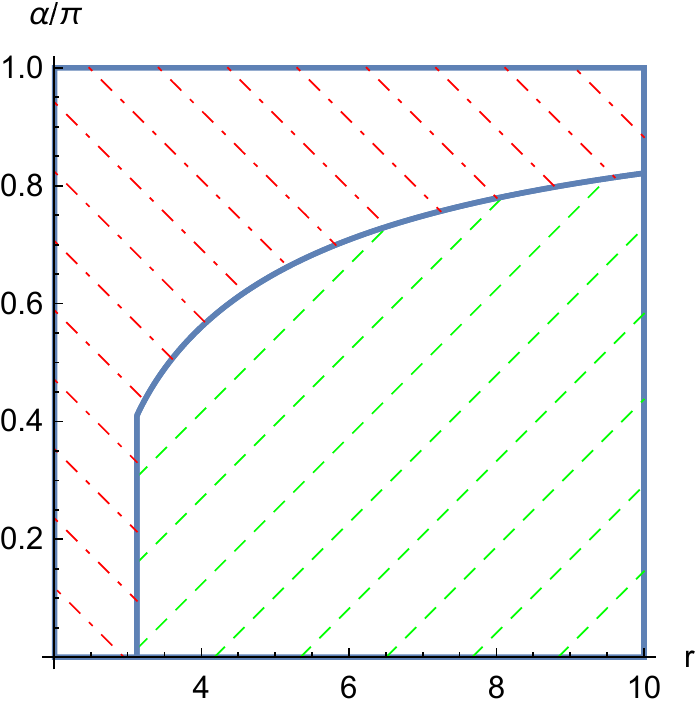}
\includegraphics[width=0.24\textwidth]{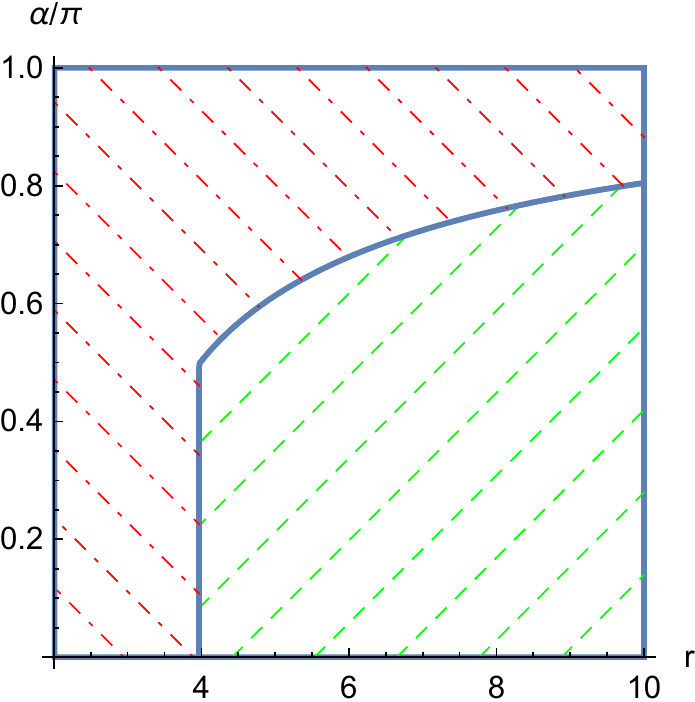}
\includegraphics[width=0.24\textwidth]{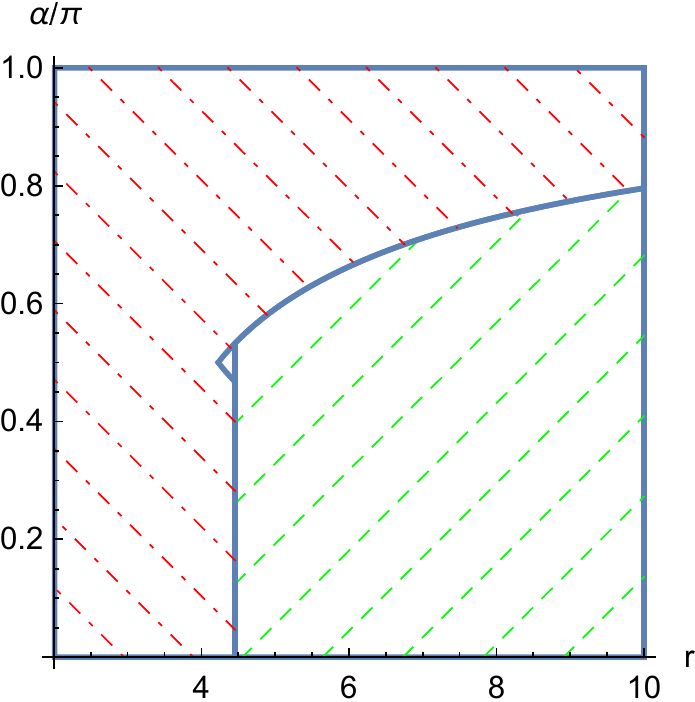}
\includegraphics[width=0.24\textwidth]{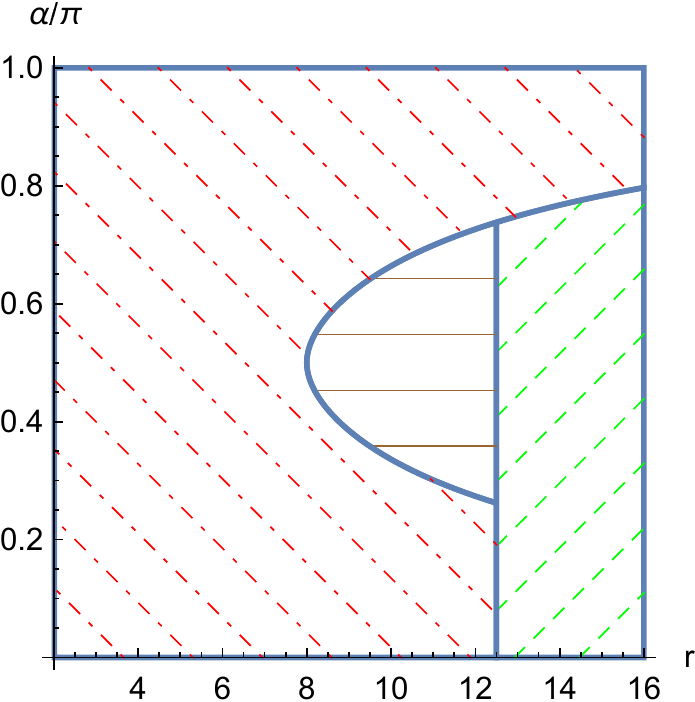}\\
(a)\hspace{3.5cm}(b)\hspace{3.5cm}(c)\hspace{3.5cm}(d)
\caption{The regions of escape (dashed green), bound (solid brown) and capture (dotdashed red) orbits for (a) $v=0.80$, (b) $v=0.71$, (c) $v=0.67$ and (d) $v=0.40$  and different $r$ and $\alpha$. \label{figslice}}
\end{figure}
\ec

To see more clearly the evolvement of various regions as velocity changes, in Fig. \refer{figslice} we show the escape, bound and caption regions in the 2-dimensional parameter space of $(r,\alpha)$ for some typical velocities from $v=0.8$ down to $v=0.4$. In each plot, the upper boundary separating the regions of capture and escape orbits, and the entire boundary (if any) separating the regions of bound and capture orbits are given by $\sin\alpha=\sin\alpha_e$ for that specific $v$. The lower boundary separating the regions of capture or bound orbits from the region of escape orbits is given by $r=2/v^2$. Similar to what was observed in Fig. \ref{figseperate}, it is seen from Fig. \ref{figslice} (a) and (b) that, for particles with velocity larger than $1/\sqrt{2}\approx 0.707$, there exists only escape or capture orbits but no bound ones. While as velocity decreases to values lower than $1/\sqrt{2}$, a region of bound orbits starts to appear and grows as velocity decreases (see Fig. \ref{figslice} (c) and (d)). Moreover from Fig. \ref{figslice} (d), one can see that for a fixed velocity and radius, the angles that allow bound orbits should be  between $\alpha=\alpha_e$ and $\alpha=\pi-\alpha_e$. In real local coordinate space, these two angles will form a double-napped cone structure.

\bc
\begin{figure}[htp!]
\includegraphics[width=0.13\textwidth]{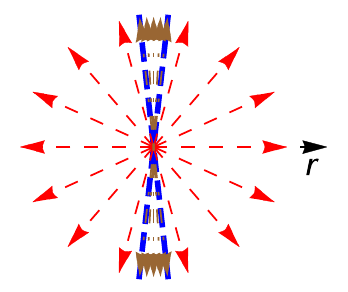}
\includegraphics[width=0.13\textwidth]{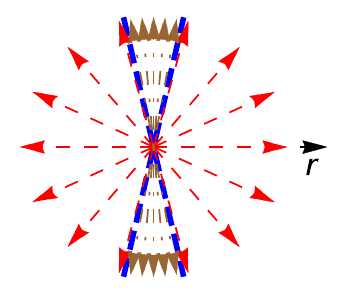}
\includegraphics[width=0.13\textwidth]{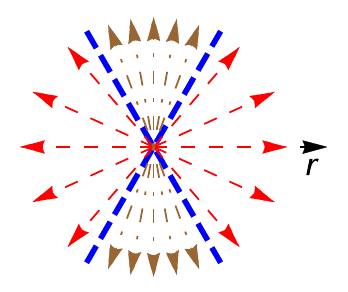}
\includegraphics[width=0.13\textwidth]{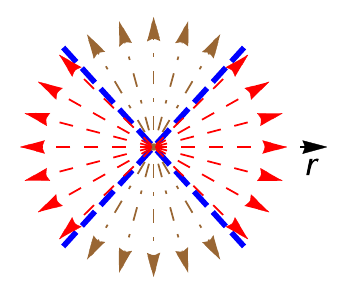}
\includegraphics[width=0.13\textwidth]{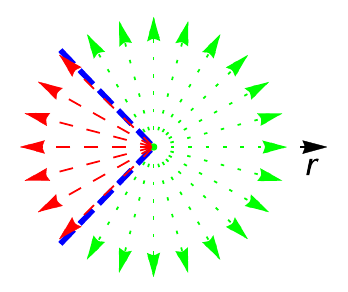}
\includegraphics[width=0.13\textwidth]{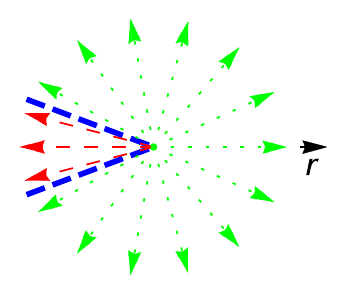}
\includegraphics[width=0.13\textwidth]{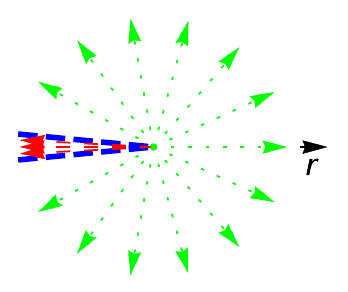}\\
(a)\hspace{2.1cm}(b)\hspace{1.9cm}(c)\hspace{1.9cm}(d)\hspace{1.9cm}(e)\hspace{1.9cm}(f)\hspace{1.9cm}(g)
\caption{The capture (dashed red arrows), bound (dotdashed brown arrows) and escape (dotted green arrows) directions for $v=0.4$ (see Fig. \ref{figslice} (d)) but different radius. From (a) to (g) the radii are respectively 8.1, 8.4, 10.0, 12.2, 12.6, 28.0 and 108.0. The angle of the inward cone of the double-napped cones or that of the single cone, denoted by the blue dashed lines, against the radial outward direction are respectively 0.534$\pi$, 0.571$\pi$, 0.664$\pi$, 0.731$\pi$, 0.74$\pi$, 0.883$\pi$ and  0.969$\pi$ rad. \label{figfixedvchanger}}
\end{figure}
\ec

We have plotted in Fig. \ref{figfixedvchanger} the directions of escape, bound and capture orbits in the real local coordinate space  for $v=0.4$ and a few increasing radii $r$. The double-napped cone structure can be seen in Fig. \ref{figfixedvchanger} (a) to (d).
As $r$ increases from 2, this double-napped cone structure is absent until $r$ reaches a critical value
\be
r=r_{\mathrm{crit1}}=\left\{
\ba{ll}
\displaystyle 2\lb \frac{2}{v}-1\rb&\displaystyle ~ \mbox{if}~ v<\frac{c}{2},\\
\displaystyle 2+\frac{1}{v^2}&\displaystyle ~ \mbox{if}~ \frac{c}{2}\leq v<\frac{c}{\sqrt{2}},
\ea
\right.
\ee
For $v=0.4$, $r_{\mathrm{crit1}}=8$ and $\alpha_e=\pi/2$
(see Fig. \ref{figfixedvchanger} (a) in which $r=8.1$). When $r$ is greater than $r_{\mathrm{crit1}}$, there appears two opposite cones forming angles $\alpha_e$ and $\pi-\alpha_e$ respectively against the radial outward direction. Orbits with initial directions inside the cones will all be captured (labeled with red arrows) and those outside the cones will follow bound orbits (labeled with brown arrows).
As $r$ increases, one can see that $\alpha_e$ decreases while $\pi-\alpha_e$ increases.
At another critical radius
\be
r=r_{\mathrm{crit2}}=\frac{2}{v^2}
\ee
the bound orbit directions and the escape directions  within the outward cone suddenly become escape directions. For $v=0.4$, $r_{\mathrm{crit2}}=12.5$.  Only the inward cone with angle $\alpha=\pi-\alpha_e$ against the outward radial direction is left, separating the regions of escape and capture orbits (see Fig. \ref{figfixedvchanger} (d) to (e) for this change). As $r$ further increases, the escape cone angle $\pi-\alpha_e$ will also increase, approaching $\pi$ near infinite $r$ (see Fig. \ref{figfixedvchanger} (f), (g)).

\bc
\begin{figure}[htp!]
\includegraphics[width=0.23\textwidth]{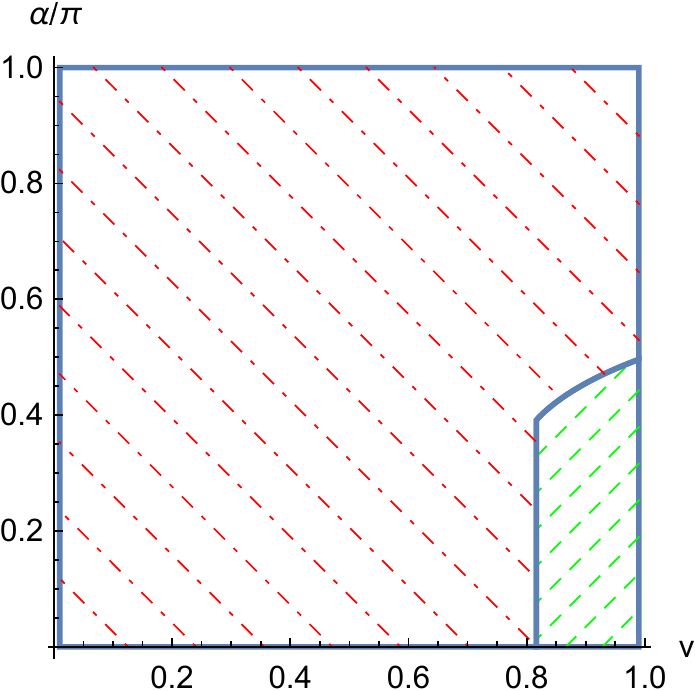}
\includegraphics[width=0.23\textwidth]{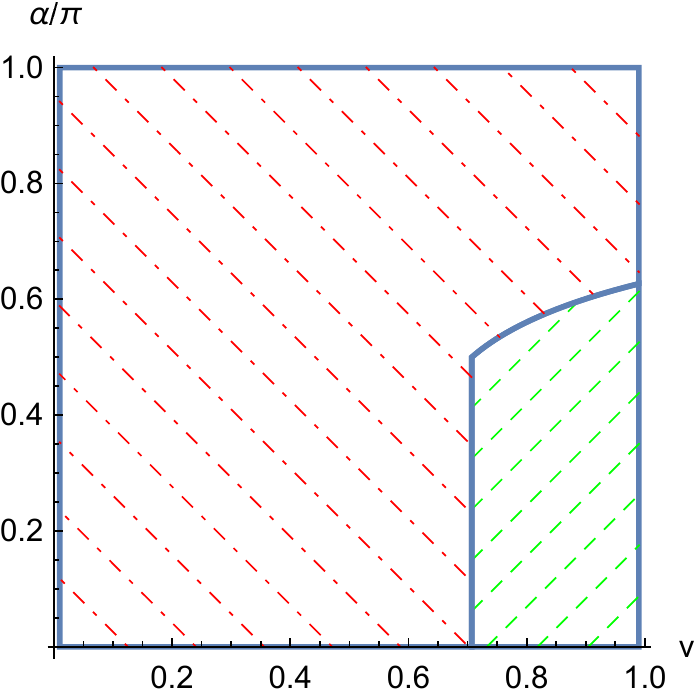}
\includegraphics[width=0.23\textwidth]{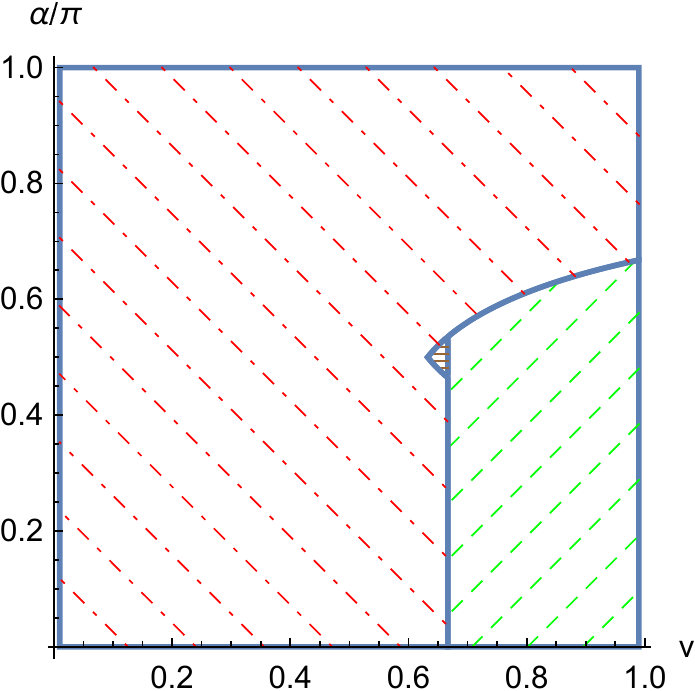}
\includegraphics[width=0.23\textwidth]{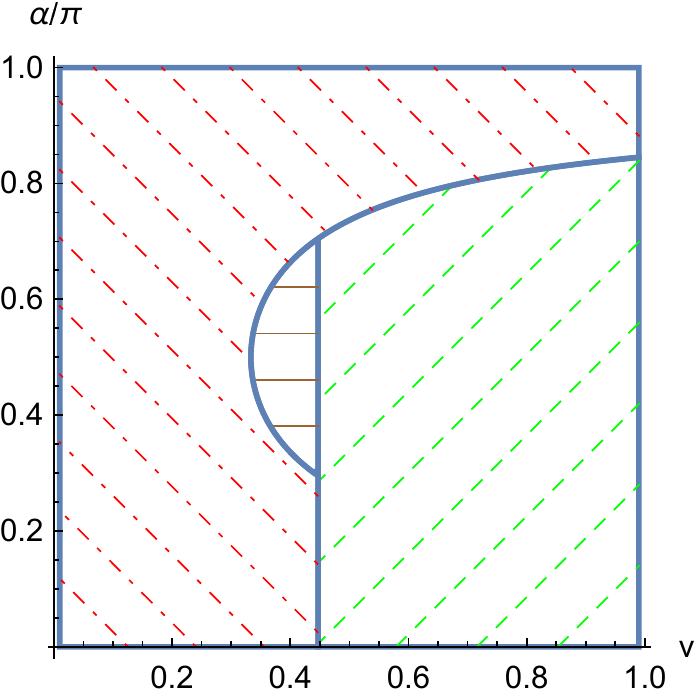}\\
(a)\hspace{3.5cm}(b)\hspace{3.5cm}(c)\hspace{3.5cm}(d)
\caption{The regions of escape (dashed green), bound (solid brown) and capture (dotdashed red) orbits for (a) $r=3$, (b) $r=4$, (c) $r=4.5$  and (d) $r=10$  and different $\alpha$ and $v$. \label{figfixedr}}
\end{figure}
\ec

To see the evolvement of various regions as $r$ changes, we plot in Fig. \refer{figfixedr} the 2-dimensional parameter space of $(v,~\alpha)$ for several fixed radii.
Similar to the case of fixed $v$ in Fig. \ref{figslice}, the boundaries separating the regions of escape orbits from regions of bound and capture orbits are also the curves $\alpha=\pi-\alpha_e$ and $r=2/v^2$. The boundary separating the regions (if any) of bound orbits and capture orbits is given by $\sin\alpha=\sin\alpha_e$. For radius $2<r<4$, one sees that there exist only escape and capture orbits but no bound orbits. The bound orbits only start to appear when radius is larger than $4$. This value agrees with the radius of particle sphere in Schwarzschild spacetime for incoming particles having almost zero velocity at infinity \cite{Jia:2015zon}.
For $r>4$, the size of the region of bound orbits grows as $r$ increases.

\bc
\begin{figure}[htp!]
\includegraphics[width=0.13\textwidth]{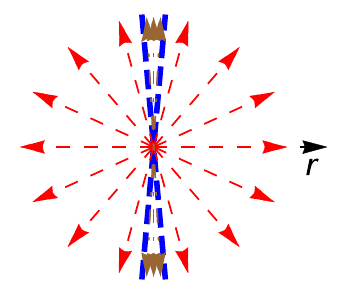}
\includegraphics[width=0.13\textwidth]{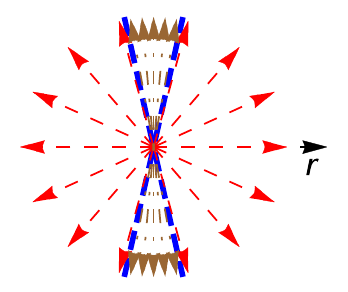}
\includegraphics[width=0.13\textwidth]{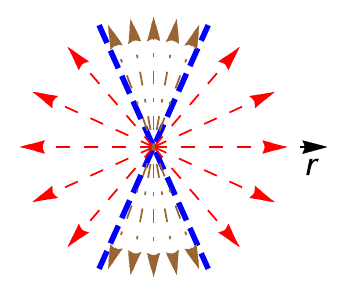}
\includegraphics[width=0.13\textwidth]{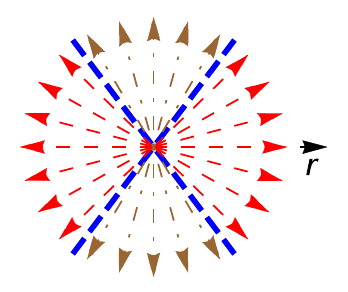}
\includegraphics[width=0.13\textwidth]{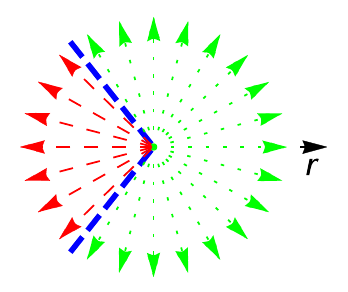}
\includegraphics[width=0.13\textwidth]{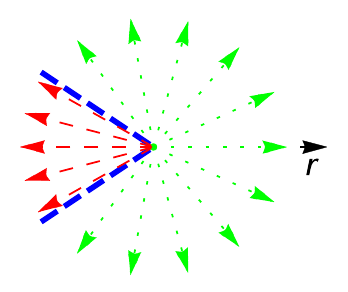}
\includegraphics[width=0.13\textwidth]{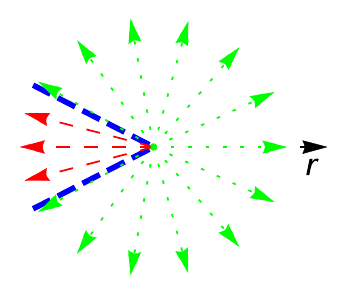}\\
(a)\hspace{1.9cm}(b)\hspace{1.9cm}(c)\hspace{1.9cm}(d)\hspace{1.9cm}(e)\hspace{1.9cm}(f)\hspace{1.9cm}(g)
\caption{The capture (dashed red arrows), bound (dotdashed brown arrows)and escape (dotted green arrows) directions for $r=10$ (see Fig. \ref{figfixedr} (d)) but different velocities. From (a) to (g) the velocities are respectively 0.335, 0.344, 0.374, 0.444, 0.454, 0.734 and 0.999.  The angle of the inward cone against the radial outward direction are respectively 0.528$\pi$, 0.569$\pi$, 0.631$\pi$, 0.703$\pi$, 0.710$\pi$, 0.810$\pi$ and 0.846$\pi$ rad. \label{figfixedrchangev}}
\end{figure}
\ec

The boundaries separating various regions in Fig. \refer{figfixedr} also correspond to cone structures in the real local coordinate system. To see the evolvement of this structure at fixed $r$ as $v$ increases, we illustrate in Fig. \ref{figfixedrchangev} the escape, bound and capture directions for several velocities for $r=10$. As the velocity increases from zero there was no escape or bound orbits in any direction until a critical velocity
\be
v=v_{\mathrm{crit1}}=\left\{\ba{ll}
\displaystyle \frac{1}{\sqrt{r-2}}&~\mbox{if}~4<r<6,\\
\displaystyle \frac{4}{2+r}&~\mbox{if}~r>6.\ea
\right.
\ee
Beyond this velocity (for $r=10$, $v_{\mathrm{crit1}}=1/3$), again the double-napped cone structure starts to appear around angle $\pi/2$ (see Fig. \ref{figfixedrchangev} (a) with $v=0.335$). The angles of the outward (or inward) cone against the radial outward direction $\alpha_e$ (or $\pi-\alpha_e$) decreases (or increases) as velocity further increases (see Fig. \ref{figfixedrchangev} (b)-(d)). At a second critical velocity
\be
v=v_{\mathrm{crit2}}=\sqrt{\frac{2}{r}},
\ee
the bound orbit directions and the capture directions within the outward cone become escape directions (see Fig. \ref{figfixedrchangev} (d)-(e) for this change). Only the inward cone structure is left, separating escape and capture geodesics.  As $v$ further increases to the speed of light, the opening angle of this cone against radial outward direction further increases until its limit value determined by $\alpha_\gamma(r)$ for photon in Eq. \eqref{photoec} (see Fig. \ref{figfixedrchangev} (f)-(g)).

\section{Implications and discussion \label{sec:impdis}}

We discuss three possible ways that the results here might be useful. The first is regarding the shadows formed by signals of nonzero mass, such as neutrinos and possibly GWs, and their implication to the properties of these particles/waves. The second and third applications are in BH accretion and spacecraft navigation respectively.

Firstly, from Fig. \ref{figfixedrchangev} (e)-(g), we see that comparing to photons, massive particles starting from the same radius will have a larger capture cone angle.
Because the geodesics we considered are reversible in time, this implies that massive particles will form a larger shadow of the same BH comparing to the shadow formed by lightrays.
For both photons and relativistic massive particles far away from the gravitational center, the Eq. \refer{alpha3def} gives a universal expression of the shadow angular size. Assuming that the galactic central BH mass is $4.1\times 10^6M_\odot$ and the distance is 8.122 kpc, its photon shadow diameter size was estimated using \eqref{photoec} to be around 51.78 $\mu$s \cite{Grenzebach:2014fha}.
While for the M87 central BH, using mass $6.2\times 10^9 M_\odot$ and distance of 16.8 Mpc, the shadow size is about 37.85 $\mu$s. On the other hand, besides electromagnetic signals, with the discovery of supernova neutrino from SN1987A \cite{Hirata:1987hu, Bionta:1987qt} and blazer TXS 0506+056 \cite{IceCube:2018dnn,IceCube:2018cha}, and the recent observation of gravitational wave (GW) signal \cite{Abbott:2016blz,Abbott:2016nmj,Abbott:2017oio,TheLIGOScientific:2017qsa} and the observation of the binary neutron star merger GW170817 and GW170817A \cite{GBM:2017lvd,Monitor:2017mdv}, it is now well known that neutrinos and GWs can also function as astrophysical messengers. Because (at least two of the three) neutrinos are massive, in principle the shadow formed by neutrino signal should be slightly larger than that of the photons. For supernova neutrinos, their typical energy is at the order of 10 MeV. Using the mass square difference in Ref. \cite{Tanabashi:2018oca} and total mass bound in Ref. \cite{Ade:2015xua}, one can estimate that neutrino velocity is at the order $v=(1-10^{-16})c$ or even larger. For relativistic signals, one can expand the angle \eqref{alpha3def} to the first order of $(1-v)$ and find the result
\be
\alpha_e(v,r)=\alpha_\gamma(r)+\frac{2\sqrt{3}}{\sqrt{(r+6)(r-2)}}(1-v)+\calco(1-v)^2. \label{eqalphaeexp} \ee
A simple calculation using Eq. \eqref{eqalphaeexp} suggests that the shadow size of neutrinos of the above velocity is very close to that of photon, i.e., 51.78 $\mu$s for galactic central BH and 37.85 $\mu$s for M87 central BH. Unfortunately, these shadow sizes for neutrino signal are beyond the angular resolution of current or near future neutrino detectors. Nevertheless, if in future the resolution did reach this limit or even better, then because different neutrino mass eigenstate will travel at different velocities, their shadow sizes will also be different. These shadow sizes therefore can be correlated with the absolute mass and mass hierarchy of neutrinos.
For GWs, its velocity difference from the speed of light has been constrained to be less than $3\times 10^{-15}c$ \cite{Monitor:2017mdv}. Therefore, using Eq. \eqref{eqalphaeexp} one can see that the shadow formed by GWs passing by the galactic or M87 center should also be around 51.78 $\mu$s or 37.85 $\mu$s respectively. Again, this value is beyond the angular resolution of current or near future GW detectors. Similar to the neutrino case however, using Eq. \eqref{alpha3def} the GW velocity can also be correlated with the shadow size.

Another circumstance where results in this work can be relevant is in the accretion of BHs. In BH accretion models, usually the falling of materials and the escape of photons and particles play crucial rules in the dynamics of the accretion. However, many of the accretion theories simply assume a simple radius or velocity of escape for particles but did not considered a detailed dependence of escape geodesic on the particle velocity, radius and escape angles \cite{Page:1974he,King:2003xf,alatalo}. Although the background spacetime in accretion models are usually more complicated than a Schwarzschild spacetime, the results here showing the detailed dependence of the escape region on the initial parameters should provide a primary step towards a better understanding of escape of particles with arbitrary velocity/angles in slightly more complicated spacetimes, e.g., Schwarzschild spacetime with (weak) magnetic field \cite{Zahrani:2013up}, Reissner-Nordstrom BHs \cite{Alonso:2007ts,Nemoto:2012cq}
and Kerr spacetime with small orbital angular momentum (with/without magnetic field) \cite{Williams:2011uz, Hussain:2014cba,Zahrani:2014rqa,Kopacek:2018lgy}.

In a somewhat different direction, the results here might be useful for the navigation of spacecraft near massive compact objects. Suppose that the spacecraft can only accelerate locally to a maximum velocity $v$ but its velocity direction can be chosen arbitrarily, and then follows a geodesic motion after acceleration, then the question is can the spacecraft escape to infinity or enter the desired bound orbit along the chosen direction and avoid being captured by the BH. In this case, the result Eq. \refer{alpha3def} and Figs. \ref{figcase1} to \ref{figfixedr}, especially the partition of the parameter space in Fig. \ref{figseperate} will provides the spacecraft necessary values of needed velocity and its direction at the given radius.

Finally, let us point out that it is straightforward to extend the current work to other spherically symmetric static spacetimes and equatorial motions in axially symmetric spacetimes. A more dramatic improvement would be to use field approach rather than the optical geodesic method to study the final state of particles. This is particularly important for the study of GWs because in this case a GW with frequency $\sim$1 Hz has wavelength comparable to size of massive BHs of $10^5 M_\odot$. If the GW frequency is 0.1 Hz or slightly below, which is well in the reach of near future GW detectors \cite{AmaroSeoane:2012je, Luo:2015ght}, then its wavelength will be larger than the $\sim 10^6M_\odot$ BH sizes and consequently its wave nature will enhance wave effect such as absorption and interference of GWs. In these situations, it would be necessary to use a field treatment to properly solve the outcome of the propagating GWs.

\begin{acknowledgments}
The authors greatly appreciate Mr. Xiankai Pang for drawing some of the figures in this work. The work of J. Liu and J. Jia was supported by the Natural Natural Science Fundation of China (No. 11504276). Nan Yang is supported by the National Natural Science Foundation of China (No. 31571797).
\end{acknowledgments}


\begin{thebibliography}{99}

\bibitem{Suzuki:1996gm} S.~Suzuki and K.~i.~Maeda,
  %``Chaos in Schwarzschild space-time: The motion of a spinning particle,''
  Phys.\ Rev.\ D {\bf 55}, 4848 (1997)
  doi:10.1103/PhysRevD.55.4848
  [gr-qc/9604020].
  %%CITATION = doi:10.1103/PhysRevD.55.4848;%%
  %106 citations counted in INSPIRE as of 08 Mar 2019

\bibitem{DOrazio:2010nbl} D.~J.~D'Orazio and P.~Saha,
  %``An analytic solution for weak-field Schwarzschild geodesics,''
  Mon.\ Not.\ Roy.\ Astron.\ Soc.\  {\bf 406}, 2787 (2010)
  doi:10.1111/j.1365-2966.2010.16879.x
  [arXiv:1003.5659 [astro-ph.GA]].
  %%CITATION = doi:10.1111/j.1365-2966.2010.16879.x;%%
  %3 citations counted in INSPIRE as of 08 Mar 2019

\bibitem{Scharf:2011ii} G.~Scharf,
  %``Schwarzschild geodesics in terms of elliptic functions and the related red shift,''
  J.\ Mod.\ Phys.\  {\bf 2}, 274 (2011)
  doi:10.4236/jmp.2011.24036
  [arXiv:1101.1207 [astro-ph.GA]].
  %%CITATION = doi:10.4236/jmp.2011.24036;%%
  %4 citations counted in INSPIRE as of 08 Mar 2019

\bibitem{Barack:2011ed} L.~Barack and N.~Sago,
  %``Beyond the geodesic approximation: conservative effects of the gravitational self-force in eccentric orbits around a Schwarzschild black hole,''
  Phys.\ Rev.\ D {\bf 83}, 084023 (2011)
  doi:10.1103/PhysRevD.83.084023
  [arXiv:1101.3331 [gr-qc]].
  %%CITATION = doi:10.1103/PhysRevD.83.084023;%%
  %79 citations counted in INSPIRE as of 08 Mar 2019

\bibitem{Gibbons:2011rh} G.~W.~Gibbons and M.~Vyska,
  %``The Application of Weierstrass elliptic functions to Schwarzschild Null Geodesics,''
  Class.\ Quant.\ Grav.\  {\bf 29}, 065016 (2012)
  doi:10.1088/0264-9381/29/6/065016
  [arXiv:1110.6508 [gr-qc]].
  %%CITATION = doi:10.1088/0264-9381/29/6/065016;%%
  %47 citations counted in INSPIRE as of 08 Mar 2019

\bibitem{DeFalco:2016yox} V.~De Falco, M.~Falanga and L.~Stella,
  %``Approximate analytical calculations of photon geodesics in the Schwarzschild metric,''
  Astron.\ Astrophys.\  {\bf 595}, A38 (2016)
  doi:10.1051/0004-6361/201629075
  [arXiv:1608.04574 [astro-ph.HE]].
  %%CITATION = doi:10.1051/0004-6361/201629075;%%
  %5 citations counted in INSPIRE as of 08 Mar 2019

\bibitem{Tejeda:2013mva} E.~Tejeda and S.~Rosswog,
  %``An accurate Newtonian description of particle motion around a Schwarzschild black hole,''
  Mon.\ Not.\ Roy.\ Astron.\ Soc.\  {\bf 433}, 1930 (2013)
  doi:10.1093/mnras/stt853
  [arXiv:1303.4068 [astro-ph.HE]].
  %%CITATION = doi:10.1093/mnras/stt853;%%
  %31 citations counted in INSPIRE as of 08 Mar 2019

\bibitem{Tsupko:2014wza} O.~Y.~Tsupko,
  %``Unbound motion of massive particles in the Schwarzschild metric: Analytical description in case of strong deflection,''
  Phys.\ Rev.\ D {\bf 89}, no. 8, 084075 (2014)
  doi:10.1103/PhysRevD.89.084075
  [arXiv:1505.06481 [gr-qc]].
  %%CITATION = doi:10.1103/PhysRevD.89.084075;%%
  %5 citations counted in INSPIRE as of 08 Mar 2019

\bibitem{Gorbatenko:2015ing} M.~V.~Gorbatenko, V.~P.~Neznamov and E.~Y.~Popov,
  %``Analysis of half-spin particle motion in static Reissner-Nordström and Schwarzschild fields,''
  J.\ Phys.\ Conf.\ Ser.\  {\bf 678}, no. 1, 012037 (2016)
  doi:10.1088/1742-6596/678/1/012037
  [arXiv:1511.05058 [gr-qc]].
  %%CITATION = doi:10.1088/1742-6596/678/1/012037;%%
  %6 citations counted in INSPIRE as of 08 Mar 2019

\bibitem{Akiyama:2019cqa}
  K.~Akiyama {\it et al.} [Event Horizon Telescope Collaboration],
  %``First M87 Event Horizon Telescope Results. I. The Shadow of the Supermassive Black Hole,''
  Astrophys.\ J.\  {\bf 875}, no. 1, L1 (2019)
  doi:10.3847/2041-8213/ab0ec7
  [arXiv:1906.11238 [astro-ph.GA]].
  %%CITATION = doi:10.3847/2041-8213/ab0ec7;%%
  %95 citations counted in INSPIRE as of 09 Jul 2019

\bibitem{Akiyama:2019eap}
  K.~Akiyama {\it et al.} [Event Horizon Telescope Collaboration],
  %``First M87 Event Horizon Telescope Results. VI. The Shadow and Mass of the Central Black Hole,''
  Astrophys.\ J.\  {\bf 875}, no. 1, L6 (2019)
  doi:10.3847/2041-8213/ab1141
  [arXiv:1906.11243 [astro-ph.GA]].
  %%CITATION = doi:10.3847/2041-8213/ab1141;%%
  %47 citations counted in INSPIRE as of 09 Jul 2019

\bibitem{hagi} Y. Hagihara, Jap. Z Astron. Geophys. {\bf 8}, 67 (1931).

\bibitem{darwin} C. Darwin, Proc. R. Soc. London A, {\bf 249}, 180 (1959).

\bibitem{mielnik} B. Mielnik and J. Plebanski, Acta Phys. Pol. {\bf 21}, 239  (1962).

\bibitem{metzner} A. W. Metzner, J. Math. Phys. {\bf 4}, 1194 (1963).

\bibitem{zeldovich} Y. B. Zeldovich I. D. Novikov, Soy. Phys. Uspekhi {\bf 8}, 522 (1965).

\bibitem{atkinson} R. Atkinson, Astron. J. {\bf 70}, 517 (1965).

\bibitem{synge1} J. L. Synge, Mon.\ Not.\ Roy.\ Astron.\ Soc. {\bf 131}, 463 (1966).

\bibitem{Chandrasekhar:1985kt} S.~Chandrasekhar,
  {\it The mathematical theory of black holes},
  OXFORD, UK: CLARENDON (1985), p 130 and p 96.
  %324 citations counted in INSPIRE as of 09 Mar 2019.

\bibitem{Perlick:2004tq} V.~Perlick,
  %``Gravitational lensing from a spacetime perspective,''
  Living Rev.\ Rel.\  {\bf 7}, 9 (2004).
  %%CITATION = 00222,7,9;%%
  %124 citations counted in INSPIRE as of 22 Mar 2019}

\bibitem{Grenzebach:2014fha} A.~Grenzebach, V.~Perlick and C.~L\"{a}mmerzahl,
  %``Photon Regions and Shadows of Kerr-Newman-NUT Black Holes with a Cosmological Constant,''
  Phys.\ Rev.\ D {\bf 89}, no. 12, 124004 (2014)
  doi:10.1103/PhysRevD.89.124004
  [arXiv:1403.5234 [gr-qc]].
  %%CITATION = doi:10.1103/PhysRevD.89.124004;%%
  %91 citations counted in INSPIRE as of 22 Mar 2019}

\bibitem{Grenzebach:2015oea} A.~Grenzebach, V.~Perlick and C.~L\"{a}mmerzahl,
  %``Photon Regions and Shadows of Accelerated Black Holes,''
  Int.\ J.\ Mod.\ Phys.\ D {\bf 24}, no. 09, 1542024 (2015)
  doi:10.1142/S0218271815420249
  [arXiv:1503.03036 [gr-qc]].
  %%CITATION = doi:10.1142/S0218271815420249;%%
  %37 citations counted in INSPIRE as of 22 Mar 2019}

\bibitem{Perlick:2015vta} V.~Perlick, O.~Y.~Tsupko and G.~S.~Bisnovatyi-Kogan,
  %``Influence of a plasma on the shadow of a spherically symmetric black hole,''
  Phys.\ Rev.\ D {\bf 92}, no. 10, 104031 (2015)
  doi:10.1103/PhysRevD.92.104031
  [arXiv:1507.04217 [gr-qc]].
  %%CITATION = doi:10.1103/PhysRevD.92.104031;%%
  %43 citations counted in INSPIRE as of 22 Mar 2019}

\bibitem{Abdujabbarov:2016hnw} A.~Abdujabbarov, M.~Amir, B.~Ahmedov and S.~G.~Ghosh,
  %``Shadow of rotating regular black holes,''
  Phys.\ Rev.\ D {\bf 93}, no. 10, 104004 (2016)
  doi:10.1103/PhysRevD.93.104004
  [arXiv:1604.03809 [gr-qc]].
  %%CITATION = doi:10.1103/PhysRevD.93.104004;%%
  %50 citations counted in INSPIRE as of 22 Mar 2019.

\bibitem{Misner:1974qy} C.~W.~Misner, K.~S.~Thorne and J.~A.~Wheeler,
  {\it Gravitation},
  San Francisco 1973, 1279p
  %381 citations counted in INSPIRE as of 09 Mar 2019

\bibitem{Jia:2015zon} X.~Liu, J.~Jia and N.~Yang,
  %``Gravitational lensing of massive particles in Schwarzschild gravity,''
  Class.\ Quant.\ Grav.\  {\bf 33}, no. 17, 175014 (2016)
  doi:10.1088/0264-9381/33/17/175014
  [arXiv:1512.04037 [gr-qc]].
  %%CITATION = doi:10.1088/0264-9381/33/17/175014;%%
  %5 citations counted in INSPIRE as of 21 Mar 2019

\bibitem{Hirata:1987hu} K.~Hirata {\it et al.} [Kamiokande-II Collaboration],
%``Observation of a Neutrino Burst from the Supernova SN 1987a,''
Phys.\ Rev.\ Lett.\ {\bf 58}, 1490 (1987).
%doi:10.1103/PhysRevLett.58.1490
%%CITATION = doi:10.1103/PhysRevLett.58.1490;%%
%1463 citations counted in INSPIRE as of 09 May 2017

\bibitem{Bionta:1987qt} R.~M.~Bionta {\it et al.},
%``Observation of a Neutrino Burst in Coincidence with Supernova SN 1987a in the Large Magellanic Cloud,''
Phys.\ Rev.\ Lett.\ {\bf 58}, 1494 (1987).
%doi:10.1103/PhysRevLett.58.1494
%%CITATION = doi:10.1103/PhysRevLett.58.1494;%%
%1268 citations counted in INSPIRE as of 09 May 2017

\bibitem{IceCube:2018dnn} M.~G.~Aartsen {\it et al.} [IceCube and Fermi-LAT and MAGIC and AGILE and ASAS-SN and HAWC and H.E.S.S. and INTEGRAL and Kanata and Kiso and Kapteyn and Liverpool Telescope and Subaru and Swift NuSTAR and VERITAS and VLA/17B-403 Collaborations],
  %``Multimessenger observations of a flaring blazar coincident with high-energy neutrino IceCube-170922A,''
  Science {\bf 361}, no. 6398, eaat1378 (2018)
  doi:10.1126/science.aat1378
  [arXiv:1807.08816 [astro-ph.HE]].
  %%CITATION = doi:10.1126/science.aat1378;%%
  %48 citations counted in INSPIRE as of 17 Jan 2019

\bibitem{IceCube:2018cha} M.~G.~Aartsen {\it et al.} [IceCube Collaboration],
  %``Neutrino emission from the direction of the blazar TXS 0506+056 prior to the IceCube-170922A alert,''
  Science {\bf 361}, no. 6398, 147 (2018)
  doi:10.1126/science.aat2890
  [arXiv:1807.08794 [astro-ph.HE]].
  %%CITATION = doi:10.1126/science.aat2890;%%
  %73 citations counted in INSPIRE as of 14 Feb 2019

\bibitem{Abbott:2016blz} B.~P.~Abbott {\it et al.} [LIGO Scientific and Virgo Collaborations],
  %``Observation of Gravitational Waves from a Binary Black Hole Merger,''
  Phys.\ Rev.\ Lett.\  {\bf 116}, no. 6, 061102 (2016)
  doi:10.1103/PhysRevLett.116.061102
  [arXiv:1602.03837 [gr-qc]].
  %%CITATION = doi:10.1103/PhysRevLett.116.061102;%%
  %3452 citations counted in INSPIRE as of 17 Jan 2019

\bibitem{Abbott:2016nmj} B.~P.~Abbott {\it et al.} [LIGO Scientific and Virgo Collaborations],
  %``GW151226: Observation of Gravitational Waves from a 22-Solar-Mass Binary Black Hole Coalescence,''
  Phys.\ Rev.\ Lett.\  {\bf 116}, no. 24, 241103 (2016)
  doi:10.1103/PhysRevLett.116.241103
  [arXiv:1606.04855 [gr-qc]].
  %%CITATION = doi:10.1103/PhysRevLett.116.241103;%%
  %1589 citations counted in INSPIRE as of 17 Jan 2019

\bibitem{Abbott:2017oio} B.~P.~Abbott {\it et al.} [LIGO Scientific and Virgo Collaborations],
  %``GW170814: A Three-Detector Observation of Gravitational Waves from a Binary Black Hole Coalescence,''
  Phys.\ Rev.\ Lett.\  {\bf 119}, no. 14, 141101 (2017)
  doi:10.1103/PhysRevLett.119.141101
  [arXiv:1709.09660 [gr-qc]].
  %%CITATION = doi:10.1103/PhysRevLett.119.141101;%%
  %716 citations counted in INSPIRE as of 17 Jan 2019

\bibitem{TheLIGOScientific:2017qsa} B.~P.~Abbott {\it et al.} [LIGO Scientific and Virgo Collaborations],
  %``GW170817: Observation of Gravitational Waves from a Binary Neutron Star Inspiral,''
  Phys.\ Rev.\ Lett.\  {\bf 119}, no. 16, 161101 (2017)
  doi:10.1103/PhysRevLett.119.161101
  [arXiv:1710.05832 [gr-qc]].
  %%CITATION = doi:10.1103/PhysRevLett.119.161101;%%
  %1533 citations counted in INSPIRE as of 17 Jan 2019

\bibitem{GBM:2017lvd} B.~P.~Abbott {\it et al.} [LIGO Scientific and Virgo and Fermi GBM and INTEGRAL and IceCube and IPN and Insight-Hxmt and ANTARES and Swift and Dark Energy Camera GW-EM and DES and DLT40 and GRAWITA and Fermi-LAT and ATCA and ASKAP and OzGrav and DWF (Deeper Wider Faster Program) and AST3 and CAASTRO and VINROUGE and MASTER and J-GEM and GROWTH and JAGWAR and CaltechNRAO and TTU-NRAO and NuSTAR and Pan-STARRS and KU and Nordic Optical Telescope and ePESSTO and GROND and Texas Tech University and TOROS and BOOTES and MWA and CALET and IKI-GW Follow-up and H.E.S.S. and LOFAR and LWA and HAWC and Pierre Auger and ALMA and Pi of Sky and DFN and ATLAS Telescopes and High Time Resolution Universe Survey and RIMAS and RATIR and SKA South Africa/MeerKAT Collaborations and AstroSat Cadmium Zinc Telluride Imager Team and AGILE Team and 1M2H Team and Las Cumbres Observatory Group and MAXI Team and TZAC Consortium and SALT Group and Euro VLBI Team and Chandra Team at McGill University],
  %``Multi-messenger Observations of a Binary Neutron Star Merger,''
  Astrophys.\ J.\  {\bf 848}, no. 2, L12 (2017)
  doi:10.3847/2041-8213/aa91c9
  [arXiv:1710.05833 [astro-ph.HE]].
  %%CITATION = doi:10.3847/2041-8213/aa91c9;%%
  %575 citations counted in INSPIRE as of 17 Jan 2019

\bibitem{Monitor:2017mdv} B.~P.~Abbott {\it et al.} [LIGO Scientific and Virgo and Fermi-GBM and INTEGRAL Collaborations],
  %``Gravitational Waves and Gamma-rays from a Binary Neutron Star Merger: GW170817 and GRB 170817A,''
  Astrophys.\ J.\  {\bf 848}, no. 2, L13 (2017)
  doi:10.3847/2041-8213/aa920c
  [arXiv:1710.05834 [astro-ph.HE]].
  %%CITATION = doi:10.3847/2041-8213/aa920c;%%
  %635 citations counted in INSPIRE as of 20 Mar 2019

\bibitem{Tanabashi:2018oca} M.~Tanabashi {\it et al.} [Particle Data Group],
  %``Review of Particle Physics,''
  Phys.\ Rev.\ D {\bf 98}, no. 3, 030001 (2018).
  doi:10.1103/PhysRevD.98.030001
  %%CITATION = doi:10.1103/PhysRevD.98.030001;%%
  %1002 citations counted in INSPIRE as of 14 Feb 2019

\bibitem{Ade:2015xua} P.~A.~R.~Ade {\it et al.} [Planck Collaboration],
  %``Planck 2015 results. XIII. Cosmological parameters,''
  Astron.\ Astrophys.\  {\bf 594}, A13 (2016)
  doi:10.1051/0004-6361/201525830
  [arXiv:1502.01589 [astro-ph.CO]].
  %%CITATION = doi:10.1051/0004-6361/201525830;%%
  %6721 citations counted in INSPIRE as of 14 Feb 2019

\bibitem{Page:1974he} D.~N.~Page and K.~S.~Thorne,
  %``Disk-Accretion onto a Black Hole. Time-Averaged Structure of Accretion Disk,''
  Astrophys.\ J.\  {\bf 191}, 499 (1974).
  doi:10.1086/152990
  %%CITATION = doi:10.1086/152990;%%
  %358 citations counted in INSPIRE as of 20 Mar 2019

\bibitem{King:2003xf} A.~R.~King and K.~A.~Pounds,
  %``Black hole winds,''
  Mon.\ Not.\ Roy.\ Astron.\ Soc.\  {\bf 345}, 657 (2003)
  doi:10.1046/j.1365-8711.2003.06980.x
  [astro-ph/0305541].
  %%CITATION = doi:10.1046/j.1365-8711.2003.06980.x;%%
  %243 citations counted in INSPIRE as of 20 Mar 2019

\bibitem{alatalo} K. Alatalo,
%Escape, accretion, or star formation? The competing depleters of gas in the quasar Markarian 231.
Astrophy. J. Lett. {\bf 801}, L17  (2015).

\bibitem{Zahrani:2013up} A.~M.~A.~Zahrani, V.~P.~Frolov and A.~A.~Shoom,
  %``Critical escape velocity for a charged particle moving around a weakly magnetized Schwarzschild black hole,''
  Phys.\ Rev.\ D {\bf 87}, no. 8, 084043 (2013)
  doi:10.1103/PhysRevD.87.084043
  [arXiv:1301.4633 [gr-qc]].
  %%CITATION = doi:10.1103/PhysRevD.87.084043;%%
  %35 citations counted in INSPIRE as of 20 Mar 2019

\bibitem{Alonso:2007ts} D.~Alonso, A.~Ruiz and M.~Sanchez-Hernandez,
  %``Escape of photons from two fixed extreme Reissner-Nordstrom black holes,''
  Phys.\ Rev.\ D {\bf 78}, 104024 (2008)
  doi:10.1103/PhysRevD.78.104024
  [gr-qc/0701052].
  %%CITATION = doi:10.1103/PhysRevD.78.104024;%%
  %5 citations counted in INSPIRE as of 20 Mar 2019

\bibitem{Nemoto:2012cq} H.~Nemoto, U.~Miyamoto, T.~Harada and T.~Kokubu,
  %``Escape of superheavy and highly energetic particles produced by particle collisions near maximally charged black holes,''
  Phys.\ Rev.\ D {\bf 87}, no. 12, 127502 (2013)
  doi:10.1103/PhysRevD.87.127502
  [arXiv:1212.6701 [gr-qc]].
  %%CITATION = doi:10.1103/PhysRevD.87.127502;%%
  %12 citations counted in INSPIRE as of 20 Mar 2019

\bibitem{Williams:2011uz} A.~J.~Williams,
  %``Numerical estimation of the escaping flux of massless particles created in collisions around a Kerr black hole,''
  Phys.\ Rev.\ D {\bf 83}, 123004 (2011)
  doi:10.1103/PhysRevD.83.123004
  [arXiv:1101.4819 [astro-ph.CO]].
  %%CITATION = doi:10.1103/PhysRevD.83.123004;%%
  %35 citations counted in INSPIRE as of 20 Mar 2019

\bibitem{Hussain:2014cba} S.~Hussain, I.~Hussain and M.~Jamil,
  %``Dynamics of a Charged Particle Around a Slowly Rotating Kerr Black Hole Immersed in Magnetic Field,''
  Eur.\ Phys.\ J.\ C {\bf 74}, no. 99, 3210 (2014)
  doi:10.1140/epjc/s10052-014-3210-y
  [arXiv:1402.2731 [gr-qc]].
  %%CITATION = doi:10.1140/epjc/s10052-014-3210-y;%%
  %19 citations counted in INSPIRE as of 20 Mar 2019

\bibitem{Zahrani:2014rqa} A.~M.~Al Zahrani,
  %``Escape of Charged Particles Moving around a Weakly Magnetized Kerr Black Hole,''
  Phys.\ Rev.\ D {\bf 90}, no. 4, 044012 (2014)
  doi:10.1103/PhysRevD.90.044012
  [arXiv:1407.7069 [gr-qc]].
  %%CITATION = doi:10.1103/PhysRevD.90.044012;%%
  %8 citations counted in INSPIRE as of 20 Mar 2019

\bibitem{Kopacek:2018lgy} O.~Kop\'{a}\u{c}ek and V.~Karas,
  %``Near-horizon structure of escape zones of electrically charged particles around weakly magnetized rotating black hole,''
  Astrophys.\ J.\  {\bf 853}, no. 1, 53 (2018)
  doi:10.3847/1538-4357/aaa45f
  [arXiv:1801.01576 [astro-ph.HE]].
  %%CITATION = doi:10.3847/1538-4357/aaa45f;%%
  %2 citations counted in INSPIRE as of 20 Mar 2019

\bibitem{AmaroSeoane:2012je} P.~Amaro-Seoane {\it et al.},
  %``Low-frequency gravitational-wave science with eLISA/NGO,''
  Class.\ Quant.\ Grav.\  {\bf 29}, 124016 (2012)
  doi:10.1088/0264-9381/29/12/124016
  [arXiv:1202.0839 [gr-qc]].
  %%CITATION = doi:10.1088/0264-9381/29/12/124016;%%
  %275 citations counted in INSPIRE as of 22 Mar 2019

\bibitem{Luo:2015ght} J.~Luo {\it et al.} [TianQin Collaboration],
  %``TianQin: a space-borne gravitational wave detector,''
  Class.\ Quant.\ Grav.\  {\bf 33}, no. 3, 035010 (2016)
  doi:10.1088/0264-9381/33/3/035010
  [arXiv:1512.02076 [astro-ph.IM]].
  %%CITATION = doi:10.1088/0264-9381/33/3/035010;%%
  %81 citations counted in INSPIRE as of 22 Mar 2019

\end{thebibliography}
\end{document}